\documentclass[showpacs,amsmath,amssymb,superscriptaddress,nofootinbib,aps,twocolumn]{revtex4}
\usepackage{multirow}
\usepackage{bigstrut}
\usepackage{amssymb}
\usepackage{amsmath}
\usepackage{graphicx}
\usepackage{dcolumn}
\usepackage{bm}
\usepackage{CJK}
\usepackage{relsize}
\usepackage{cancel}
\usepackage{color}

\begin{document}
\title{Impurity-driven two-dimensional  spin relaxation induced by  intervalley spin-flip scattering in silicon}

\author{Yang Song}\email{ysong128@umd.edu}\affiliation{Condensed Matter Theory Center, Department of Physics, University of Maryland, College Park, MD 20742, USA}\affiliation{Joint Quantum Institute, University of Maryland, College Park, MD 20742, USA}
\author{S.~Das Sarma}
\affiliation{Condensed Matter Theory Center, Department of Physics, University of Maryland, College Park, MD 20742, USA}\affiliation{Joint Quantum Institute, University of Maryland, College Park, MD 20742, USA}

\begin{abstract}
Through the theoretical study of electron spin lifetime in the two-dimensional electron gas (2DEG) confined near the surface of doped Si, we highlight a dominant spin relaxation mechanism induced by the impurity central-cell potential near an interface via intervalley electron scattering. At low temperatures and with modest doping, this Yafet spin flip mechanism can become more important than the D'yakonov-Perel' spin relaxation arising from the structural Rashba  or Dresselhaus spin-orbit coupling field. As the leading-order impurity-induced spin flip happens only between two non-opposite valleys in Si,  2DEG systems in Si MOSFETs or SiGe heterostructures are a natural platform to test and utilize this spin relaxation mechanism due to the valley splitting near the interface and the tunability by electrical gating or applied stress. Our proposed new spin relaxation mechanism may explain a part of the spin relaxation contribution to Si-based 2DEG systems,  and should have spintronic applications in Si-based devices.
\end{abstract}
\maketitle

\section{Introduction}

Silicon takes the unique position in both the conventional main-stream electronic industry (i.e. C-MOS) and the emerging fields of quantum information science and technology such as spintronics \cite{Zutic_RMP04, Jansen_NatMat12, Sverdlov_PR15} and quantum computation \cite{Kane_Nat98, Zwanenburg_RMP13}. Its strength derives from the matured capacity of extreme high purity  and low cost material growth, and perhaps more crucially, the orders of magnitude tunability in electrical conductivity enabled by doping and gating. Silicon (Si) has continued to reveal another crucial property, that is, its long spin lifetime due to the relatively small atomic spin-orbit coupling (SOC), bulk inversion symmetry, and zero nuclear spin in the abundant isotope  $^{28\!}$Si. Two-dimensional electron gas (2DEG) occupying the few lowest quantized 2D subbands of various Si surfaces and quantum wells has long become an important playground for fundamental science \cite{Ando_RMP82, Klitzing_RMP86} and more recently, quantum computing qubit platforms through gate-defined or donor-defined quantum dots \cite{Zwanenburg_RMP13}. In particular, the long spin lifetime and the ability to control the confined electrons through externally applied electrical voltage (i.e. fast gates) near the Si surface are the main drivers of the great interest and activity on Si-based spintronics and quantum computing architectures.

With respect to spin relaxation, there are some key differences between 3D and 2D Si systems worth emphasizing right at the outset. In particular, the confining potential at the interface, where the 2DEG resides,  may  break the inversion symmetry of the Si crystal. This generally results in spin splitting in the band structure and induces an effective momentum-dependent magnetic field for conduction electrons \cite{Bychkov_JPC84},  often referred to as the Rashba field arising purely from the structural asymmetry in real space. In addition, quantum wells with odd number of Si layers or broken rotoinversion symmetry at the Si-Ge interface \cite{Golub_PRB04, Nestoklon_PRB06, Nestoklon_PRB08, Prada_NJP11} induce generalized Dresselhaus field \cite{Dresselhaus_PR55, Krebs_PRL96}.

Such structural SOC effects obviously are not present inside the 3D bulk Si and can exist only in the 2DEG. When electrons undergo momentum scattering (e.g. by impurities or phonons) in the presence of Rashba/Dresselhaus effect, their spins precess randomly over time and relax \cite{Dyakonov_SPSS72}. This D'yakonov-Perel' (DP) process has been the only main spin relaxation mechanism studied so far in Si 2DEG \cite{Wilamowski_PRB02, Wilamowski_PRB04, Tahan_PRB05, Pershin_NJP07, Dugaev_PRB09}. This has led to the general belief, questioned in the current work, that the DP mechanism is the only spin relaxing mechanism in Si 2DEG that needs to be considered theoretically.

In this work, we bring in a fundamentally different spin-relaxation mechanism, which may gradually dominate over the DP mechanism with increasing impurity densities or doping. This new impurity-induced spin relaxation mechanism  does not rely on the effective Rashba/Dresselhaus magnetic field between scattering events, but rather flips spins right at the scattering events, through the contact spin-orbit interaction  at the impurity core. Therefore our present mechanism can be termed a \textit{Yafet} process \cite{Yafet_SSP63}.

While the scattering is driven by impurities in both mechanisms at low temperatures, the difference is that scattering serves to interrupt the spin precession in the DP process whereas it facilitates spin flip in the Yafet process. As a result, instead of weakening with higher impurity density or lower mobility as in the DP spin relaxation, our mechanism grows stronger with increasing (decreasing) impurity density (mobility),  and therefore can be distinguished experimentally from the DP process.

We believe that some contributions of this new impurity-induced Yafet process to Si 2D spin relaxation may have already been detected experimentally as we discuss later in this article.  We note that in 3D Si, our process is already known to be the experimentally dominant electronic spin relaxation mechanism in high doping situations \cite{Song_PRL14}.

The other aspect we highlight is the tunability of spin lifetime. While the charge transistor builds on the tunable conductivity (i.e. tunable carrier momentum relaxation time), it is desirable that the spin relaxation time can be controlled as well for spintronic applications. As we will show in detail,  the leading-order spin flip occurs only during intervalley scattering and between two non-opposite valleys among the six Si conduction valleys (so called ``$f$-process'' \cite{Yu_Cardona_Book}), whereas it vanishes during intravalley scattering or intervalley scattering between two opposite valleys (``$g$-process''). For various 2DEG plane orientations relative to the Si crystallographic direction (e.g. 100, 110, 111 or arbitrary orientation), it is well known that the  resulting 2D electronic ground states have different valley configurations, with the ground state valley degeneracy varying from 1 to 6 depending on surface orientations and details \cite{Dorda_PRB78, Ando_RMP82}. As such, the spin   lifetime determined by this mechanism in 2DEG will be distinguishable just owing to different plane orientations of the 2D system, producing substantial anisotropy in the 2D spin relaxation. Moreover, the tunability of the relative valley energies by stress and especially, by gate voltage in Si MOSFETs, can enable fast on-chip  spin lifetime control. Spin-orientation dependence of the spin relaxation, absent for the charge mobility, can also be similarly controlled. Since valleys do not play a central role in the DP process, which is governed entirely by the structural asymmetry, such orientation or gate voltage dependence of spin relaxation is qualitatively different in the DP mechanism \cite{Wilamowski_PRB04, Tahan_PRB05}. This difference  can also distinguish our proposed mechanism from the DP mechanism with respect to Si 2D spin relaxation.

In 3D bulk Si, as we mentioned this novel Yafet process has been shown to be the dominant spin relaxation mechanism when the scattering is caused by donor impurities \cite{Song_PRL14}. It is caused by the spin-dependent interaction with the impurity core, and is far more important than the spin flip during intravalley scattering by the long-range Coulomb interaction, or during intervalley scattering by the spin-independent part of the impurity core potential, neither of which exhibits the empirically strong donor dependence  \cite{Ue_PRB71,Quirt_PRB72,Pifer_PRB75,Ochiai_PSS78,Zarifis_PRB98}.  Our goal here is to introduce this important mechanism into the Si 2DEG, build up its primary trend qualitatively and quantitatively, and discuss its experimental relevance and applications. Since the DP process and our process are completely independent spin relaxation mechanisms, generically both should be present in Si 2DEG, and their relative quantitative importance will depend on all the details of the specific system and samples being studied.

We briefly discuss the scenarios where our proposed mechanism can be of importance and utilized. The first apparent criterion is low temperature and moderate-to-high impurity density, so that phonon-driven spin relaxation is relatively weak. Conversely, our mechanism is most likely overshadowed by the DP spin relaxation process in intrinsic to low-doped 2DEG systems, typical for Si/SiGe quantum wells and other modulation-doped heterostructures where interface impurity scattering is relatively weak \cite{Wilamowski_PRB02,Tyryshkin_PRL05}, except for symmetrically designed wells \cite{Sherman_PRB03, Golub_PRB04} where the structural asymmetry can be reduced. It can be easily shown that DP spin relaxation alone leads to rapidly diverging spin lifetime once the mobility $\mu$ is lowered to a few m$^2$/Vs \cite{Tahan_PRB05}.  Also, as the present mechanism relies on the SOC  between the free electron and the impurity core, for a given impurity density its quantitative effect is ranked according to the sign of impurity charges: positively charge impurities $>$ neutral impurities $\gg$ negatively charged impurities, the last of which repel the electrons and render little central-cell correction \cite{Ralph_PRB75, El-Ghanem_JPC80, Ridley_book}. As such, both n-type SiGe quantum well and accumulation layer in n-type MOSFET are good candidates for studying our mechanism. In low-mobility Si MOSFET samples, however, inversion as well as accumulation layers can be both relevant due to the dominant interface oxide charges \cite{Hartstein_SurfSci76,Ando_RMP82}, which could scatter carriers strongly leading to a strengthening of our mechanism. Noticeably, this mechanism is more effective in 2D than in 3D Si, as the former retains ionized donors under most relevant experimental conditions \cite{DasSarma_PRB13}.

In Sec.~\ref{sec:theoretical formulation} we develop our basic theory of 2D spin relaxation, obtaining detailed results for the relaxation time for different surface orientations and applied external stress in Sec.~\ref{sec:detailed_results_for_2DEG}. Section~\ref{sec:experimental_implications} is devoted to a discussion of our results in the context of experimental implications and the existing 2D spin relaxation experiments.   We conclude in Sec.~\ref{sec:summary} with a summary and an outlook. The intervalley scattering physics and its relevant symmetry analysis is reviewed in the Appendix.

\section{Theoretical formulation}\label{sec:theoretical formulation}

Our spin relaxation mechanism arises directly from the impurity SOC, in contrast with the structural SOC effects that emerge from a combination of atomic SOC and broken structural symmetry. The electron spins flip upon scattering.   Moreover, the spin flip is governed not by the spin mixing in the conduction electron states but by the SOC of the scattering potential. Our approach to formulate the scattering of electron states in the 2DEG subbands by an impurity potential is to take doubly-applied effective mass approximation (EMA) \cite{Wannier_PR37, Luttinger_PR55, Bir_Pikus_Book}. One of the EMA is conventionally applied with the envelope functions of subband states confined in a quantum well or near the surface \cite{Ando_RMP82, Burt_JPCM92} (as discussed in more detail below, we do not consider the opposite-valley coupling due to interface in our leading-order theory). Assuming the scattering matrix element for conduction states in a 3D bulk Si of volume $V$  is $U^{3d}_{v_1,\mathbf{s};v_2,\!-\mathbf{s}}$ between valley $v_1$, spin $\mathbf{s}$ and valley $v_2$, spin $-\mathbf{s}$, the EMA connects it to that of the 2DEG with an area $S$,
\begin{eqnarray}\label{eq:U_2d0}
U^{2d}_{\!v_1,\!n_1; v_2,\!n_2}\!(z,\mathbf{s})=
 \frac{\xi_{v_1,\!n_1}(z) \xi_{v_2,\!n_2}(z)}{S} V  U^{3d}_{v_1,\mathbf{s};v_2,\!-\mathbf{s}} ,
\end{eqnarray}
for a given impurity located at $z$ along the width direction of the 2DEG,  where $\xi_{v,n}(z)$ is the envelope function in valley $v$ and quantized 2D subband $n$, and normalized $\int dz \xi_{v,n}\xi_{v,m}=\delta_{n,m}$. In the following, we first elaborate the physics of the spin-flip matrix element $U^{3d}_{v_1,\mathbf{s};v_2,\!-\mathbf{s}}$, where another use of the EMA is crucial to relate a scattering problem with a donor-state problem. Then we study in detail the different specific confinements and resulting subbands.

The bulk spin-flip scattering is treated rigorously in terms of the general symmetry of the impurity potential \cite{Song_PRL14}. Since the initial and final conduction states are the eigenstates of the bulk Si (one-body) Hamiltonian $V_0$, the scattering potential is the difference between the substitutional impurity and the original Si atom, $U=V_{\rm imp}-V_{\rm Si}$, and breaks the $O^7_h$ space symmetry of the diamond lattice structure. Without going into the details of $U$ (including the screened Coulomb potential and short-ranged central-cell correction), $U$ obeys the tetrahedral $T_d$ point group symmetry \cite{Kohn_SSP57} and one can derive the matrix element form ($ U^{3d}_{v_1,\mathbf{s};v_2,\!-\mathbf{s}}= \langle \psi_{v_2,-\mathbf{s}}| U| \psi_{v_1,\mathbf{s}}\rangle$) with the  correct dependence on valleys and spin orientation of the involved conduction states \cite{Song_PRL14}. We summarize the relevant intervalley scattering in bulk Si and its symmetry analysis in the Appendix. Between conduction states at the valley centers, it turns out that spin flip survives only in intervalley $f$-process scattering. Its counterparts in intravlley and  intervalley $g$-process scattering are forbidden by the $C_2$ rotation symmetry of the $T_d$ group and the time reversal symmetry, respectively. We denote the bare spin angular dependence of $U^{3d}_{v_1,\mathbf{s};v_2,-\mathbf{s}}$ without the quantitative prefactor as $\hat{U}_{v_1,\mathbf{s};v_2,-\mathbf{s}}$, and the expression between $+x$ and $+y$ valleys for arbitrary spin orientation $\mathbf{s}=(\sin\theta \cos\phi, \sin\theta \sin\phi,\cos\theta)\equiv (s_x,s_y, s_z)$ reads \cite{Song_PRL14},
\begin{eqnarray}\label{eq:U_s_free}
\hat{U}_{+x,\!\mathbf{s}; +y,\!-\mathbf{s}}
\!\!&=&\!\! \frac{i}{6}\sin\theta e^{i\phi}
\!+\! \frac{\eta(1\!-\!i)}{2\sqrt{3}}(\cos^2\frac{\theta}{2}\!-\!i\sin^2\frac{\theta}{2} e^{2i\phi})
\nonumber\\
\!\!&\equiv&\!\! \frac{is_x\!- \!s_y}{6}
+ \frac{\eta(1\!-\!i)}{4\sqrt{3}}\left(\!1\!+\!s_z\!-\!  \frac{i(s_x\!+\!is_y)^2}{1+s_z}\!\right),\quad
\end{eqnarray}
where the dimensionless constant $\eta$ is the ratio between the two symmetry-allowed terms (in particular, from the $\bar{F}$-symmetry states of the $T_d$ group; see Appendix for details). This leads to the anisotropic dependence of spin relaxation on spin orientation.

In order to determine the magnitude of the prefactor in $U^{3d}$, we make an important connection between it and the spin-split spectrum of the localized impurity states, using the essence of  EMA. By comparing the scattering problem and the localized eigenenergy problem of the same impurity, one can realize that the potential is exactly the same for the two problems and the only difference between the localized state and the conduction state comes from the envelope function in the former due to the Coulomb confinement. As a result, the prefactor in $U^{3d}$ can be related to the spin splitting $\Delta_{\rm so}$ of the bound impurity states such that $U^{3d}_{v_1,\mathbf{s};v_2,-\mathbf{s}}=(\pi a^3_B/V)\Delta_{\rm so} \hat{U}_{v_1,\mathbf{s};v_2,-\mathbf{s}}$, where $a_B$ is the impurity Bohr radius and $V$ the bulk volume, an EMA effect not too different from that of Eq.~(\ref{eq:U_2d0}) applied for the 2DEG confinement.  When the experimental spectrum is available for $\Delta_{\rm so}$, such as those in group V donors \cite{Aggrawal_PR65, Castner_PR67}, this  method is most efficient and also likely more accurate than numerical calculations that may miss part of the microscopic contributions. The ratio constant $\eta$ is estimated to be about 2 from spin relaxation data in highly doped n-type Si \cite{Song_PRL14}. For other types of substitutional impurities $\eta$ is expected to have a  value of the order of unity. In principle, an estimate of $\eta$ can be obtained by first principles calculations which are out of scope for the current work. Again it is much preferable by empirically comparing with experiments since a precise quantitative calculation of $\eta$ is essentially impossible theoretically, particularly in the context of spin relaxation in the 2DEG. As discussed in the Introduction, in general, the SOC scattering strength depends on the type of impurities being considered, and would in general be smaller for neutral and negatively charged impurities.

$\hat{U}_{+x,\mathbf{s};+y,-\mathbf{s}}$ in Eq.~(\ref{eq:U_s_free}) describes the leading-order in wavevector spin-flip matrix element in one of all the 24 paths of the $f$-process scattering among six Si conduction valleys. In this work we group $\hat{U}_{v_1,\mathbf{s}; v_2,-\mathbf{s}}$  into 12 time-reversal (TR) related pairs, and then connect them to $\hat{U}_{+x,\mathbf{s}; +y,-\mathbf{s}}$  by specific spatial symmetry operations in the $T_d$ group:
\begin{eqnarray}\label{eq:U_s_all_f}
|\hat{U}_{v_1,\mathbf{s};v_2,-\mathbf{s}}|&\stackrel{\rm{TR}}{=} &|\hat{U}_{-v_2,\mathbf{s};-v_1,-\mathbf{s}}|,
\\
|\hat{U}_{x,\mathbf{s}; -y,-\mathbf{s}}|&\stackrel{C_{2x}}{=}& |\hat{U}_{x,\mathbf{s}'=(s_x,-s_y,-s_z);y,-\mathbf{s}'}|,
\nonumber\\
|\hat{U}_{y,\mathbf{s}; x,-\mathbf{s}}|&\stackrel{\sigma_{\mathbf{x}-\mathbf{y}}}{=}& |\hat{U}_{x,\mathbf{s}'=(-s_y,-s_x,-s_z);y,-\mathbf{s}'}|,
\nonumber\\
 |\hat{U}_{-y,\mathbf{s}; x,-\mathbf{s}}|&\stackrel{S_4}{=}& |\hat{U}_{x,\mathbf{s}'=(s_y,-s_x,s_z);y,-\mathbf{s}'}|,
\nonumber\\
|\hat{U}_{x,\mathbf{s}; z,-\mathbf{s}}|&\stackrel{\sigma_{\mathbf{y}-\mathbf{z}}}{=}& |\hat{U}_{x,\mathbf{s}'=(-s_x,-s_z,-s_y);y,-\mathbf{s}'}|,
\nonumber\\
 |\hat{U}_{x,\mathbf{s}; -z,-\mathbf{s}}|&\stackrel{\sigma_{\mathbf{y}+\mathbf{z}}}{=}& |\hat{U}_{x,\mathbf{s}'=(-s_x,s_z,s_y);y,-\mathbf{s}'}|,
\nonumber\\
 |\hat{U}_{z,\mathbf{s}; x,-\mathbf{s}}|&\stackrel{C_3}{=}& |\hat{U}_{x,\mathbf{s}'=(s_z,s_x,s_y);y,-\mathbf{s}'}|,
\nonumber\\
  |\hat{U}_{-z,\mathbf{s}; x,-\mathbf{s}}|&\stackrel{C_{3}}{=}& |\hat{U}_{x,\mathbf{s}'=(-s_z,s_x,-s_y);y,-\mathbf{s}'}|.
\nonumber\\
|\hat{U}_{z,\mathbf{s}; y,-\mathbf{s}}|&\stackrel{\sigma_{\mathbf{x}-\mathbf{z}}}{=}& |\hat{U}_{x,\mathbf{s}'=(-s_z,-s_y,-s_x); y,-\mathbf{s}'}|,
\nonumber\\
|\hat{U}_{-z,\mathbf{s}; y,-\mathbf{s}}| &\stackrel{\sigma_{\mathbf{x}+\mathbf{z}}}{=}& |\hat{U}_{x,\mathbf{s}'=(s_z,-s_y,s_x);y,-\mathbf{s}'}|,
\nonumber\\
|\hat{U}_{y,\mathbf{s}; z,-\mathbf{s}}|&\stackrel{C_3}{=}& |\hat{U}_{x,\mathbf{s}'=(s_y,s_z,s_x);y,-\mathbf{s}'}|,
\nonumber\\
 |\hat{U}_{y,\mathbf{s}; -z,-\mathbf{s}}| &\stackrel{C_{3}}{=}& |\hat{U}_{x,\mathbf{s}'=(s_y,-s_z,-s_x);y,-\mathbf{s}'}|,
\nonumber
\end{eqnarray}
where the vector subscript of  the reflection operator ($\sigma$) marks the normal direction of the reflection plane, and $C$ and $S$ denote the usual proper and improper rotations respectively with the given axes (the unspecified axis of the $C_3$ rotation is along one of the cubic body diagonals). These individual $U_{v_1,\mathbf{s}; v_2,-\mathbf{s}}$ expressions are important in evaluating spin relaxation in Si 2DEG, where not all valleys are always equally occupied and can be subsequently summed together. More specifically, the anisotropy of effective mass and strain may split the energy degeneracy of the six valleys in different ways, but always keep energy the same for the two  opposite valleys. This is true without including the small effects from SOC and short-wavelength perturbation beyond the EMA, which could induce splitting of the order of 1 meV or less \cite{Campo_SSC81,Ando_RMP82, Friesen_PRB07, Saraiva_PRB09, Saraiva_PRB11, Zwanenburg_RMP13}. This splitting effect is negligible compared with typical Fermi levels and in the context of the leading-order calculation of the spin relaxation time. We therefore do not include such small interface coupling between opposite valleys in the current work. All in all, we are always allowed to group the 24 $f$-process paths into three parts, (I) $\pm x \leftrightarrow \pm y$, (II) $\pm x \leftrightarrow \pm z$ and (III) $\pm y \leftrightarrow \pm z$, and sum $|U_{v_1,\mathbf{s}; v_2,-\mathbf{s}}|^2$ over each group which shares the same electron statistical distribution factor. Utilizing Eqs.~(\ref{eq:U_s_free}) and (\ref{eq:U_s_all_f}), we have,
\begin{eqnarray}
\sum_{8\in\; {\rm I}} |\hat{U}_{v_1,\mathbf{s}; v_2,-\mathbf{s}}|^2\!&= &\!
\frac{2}{9} [1-s_z^2 +3\eta^2 (1+s_z^2)] \equiv \mathcal{S}(s_z), \quad\label{eq:U_xy}
\\
\sum_{8\in\; \textrm{II}} |\hat{U}_{v_1,\mathbf{s}; v_2,-\mathbf{s}}|^2\!&= &\!
\mathcal{S}(s_y),  \label{eq:U_zx}
\\
\sum_{8\in\; \textrm{III}} |\hat{U}_{v_1,\mathbf{s}; v_2,-\mathbf{s}}|^2 \!&= &\!
\mathcal{S}(s_x).  \label{eq:U_zy}
\end{eqnarray}
We will see that Eqs.~(\ref{eq:U_xy})-(\ref{eq:U_zy}) directly lead to the strong spin angular dependence of the spin relaxation in the (110) and (001)-oriented 2DEGs, as well as in the (111) 2DEG under external stress in the next section.

Next we address the specific confinements and subband envelope functions  in Eq.~(\ref{eq:U_2d0}). Before doing that, we combine Eq.~(\ref{eq:U_2d0}) and $U^{3d}_{v_1,\mathbf{s};v_2,-\mathbf{s}}=(\pi a^3_B/V)\Delta_{\rm so} \hat{U}_{v_1,\mathbf{s};v_2,-\mathbf{s}}$ to give
\begin{eqnarray}\label{eq:U_2d}
U^{2d}_{\!v_1,\!n_1; v_2,\!n_2}\!(z,\mathbf{s})=
 \frac{\xi_{v_1,\!n_1}(z) \xi_{v_2,\!n_2}(z)}{S} \pi a^3_B \Delta_{\rm so} \hat{U}_{v_1,\mathbf{s};v_2,\!-\mathbf{s}} .
\end{eqnarray}
We stress that the EMA suits our problem especially well, even for the relatively narrow 2DEG: As we have shown in Ref.~\cite{Song_PRL14}, the relevant intervalley spin scattering potential comes from the core region of the impurities, evidenced by the strong dependence of the spin relaxation times on the donor species. The overall 2D confinement is much smoother than the impurity core potential whose linear dimension is much less than a lattice constant, and Eq.~(\ref{eq:U_2d}) can be safely used for most of the randomly or uniformly distributed impurities in the 2DEG. The 2DEG system is essentially of 3D nature with respect to the short-range scattering in the immediate impurity core region since the 2D confinement length scale ($\sim$ 10 nm or larger) is much larger than the atomic core size ($\sim$ 0.1 nm). In another word, the weak (as to the influence on impurity cores) symmetry-breaking potential from the 2D confinement is taken into account by the mostly slowly-varying envelope $\xi_{v,n}(z)$, but otherwise the scattering interaction and the conduction Bloch functions $\psi_{v,\mathbf{s}}$ near the impurity core region are unchanged to this order of perturbation. Thus the spin-flip selection rules are still dictated by the bulk symmetry, well retained near the impurity core region.  Under this level of approximation, we  also neglect any small change in $\Delta_{\rm so}$ and $\eta$, and in $\hat{U}_{v_1,\!\mathbf{s}; v_2,\!-\mathbf{s}}$, in going from the 3D bulk to the 2DEG.

We define an effective width $d_{v_1,n_1;v_2,n_2}$ for the scattering between $v_1,n_1$ and $v_2,n_2$ states, in terms of the envelope functions in Eq.~(\ref{eq:U_2d}),
\begin{eqnarray} \label{eq:d_rev}
\frac{1}{d_{v_1,n_1;v_2,n_2}}\equiv\int d z |\xi_{v_1,n_1}(z)\xi_{v_2,n_2}(z)|^2 ,
\end{eqnarray}
which, together with the 2DEG area $S$, yields an effective volume of the 2DEG $Sd_{v_1,n_1;v_2,n_2}$ (taking the role of $V$ in 3D bulk) for a given subband transition.
To be specific, we choose two representative confinements for the 2DEG. The first one is a square well, corresponding to the typical 2D heterostructure quantum well (such as SiGe/Si/SiGe). Focusing on the lowest few levels, we approximate the well potential as an infinite barrier for $0<z<d$ (where $d$ here is the physical well width) and obtain simple analytical solutions,
\begin{eqnarray}
\xi^{\rm sq}_{v,n}(z)= \sqrt{\frac{2}{d}} \sin \frac{(n+1) \pi z}{d},
\end{eqnarray}
where $n=0,1,2,...$ denotes various 2D confined subbands. The corresponding energies at subband bottoms are
\begin{eqnarray}\label{eq:E_n_sq}
E_{v,n}=  \frac{[\pi \hbar (n+1)]^2}{2 m_{z,v} d^2},
\end{eqnarray}
where the effective mass $m_{z,v}$ along the $z$ direction depends on the valley $v$ and the 2DEG plane orientation as we will describe in detail below.  Note that here $E_{v,n}$ is measured from the bottom of the $v$th valley, $E_v$, in the 3D bulk, as opposed to the lowest subband bottom. Different valley bottoms may  shift relatively to each other upon various stress configurations (either deliberately applied from outside or present  because of intrinsic interface strain).
Following these $\xi_{v,n}(z)$, we can obtain the effective width parameter [Eq.~(\ref{eq:d_rev})] for square wells as
\begin{eqnarray} \label{eq:d_12_sq}
d^{\rm sq}_{n_1;n_2} = \frac{d}{1+\delta_{n_1,n_2}/2},
\end{eqnarray}
which is independent of the involved valleys.

\begin{figure}[!htbp]
\centering
\includegraphics[width=8.5cm]{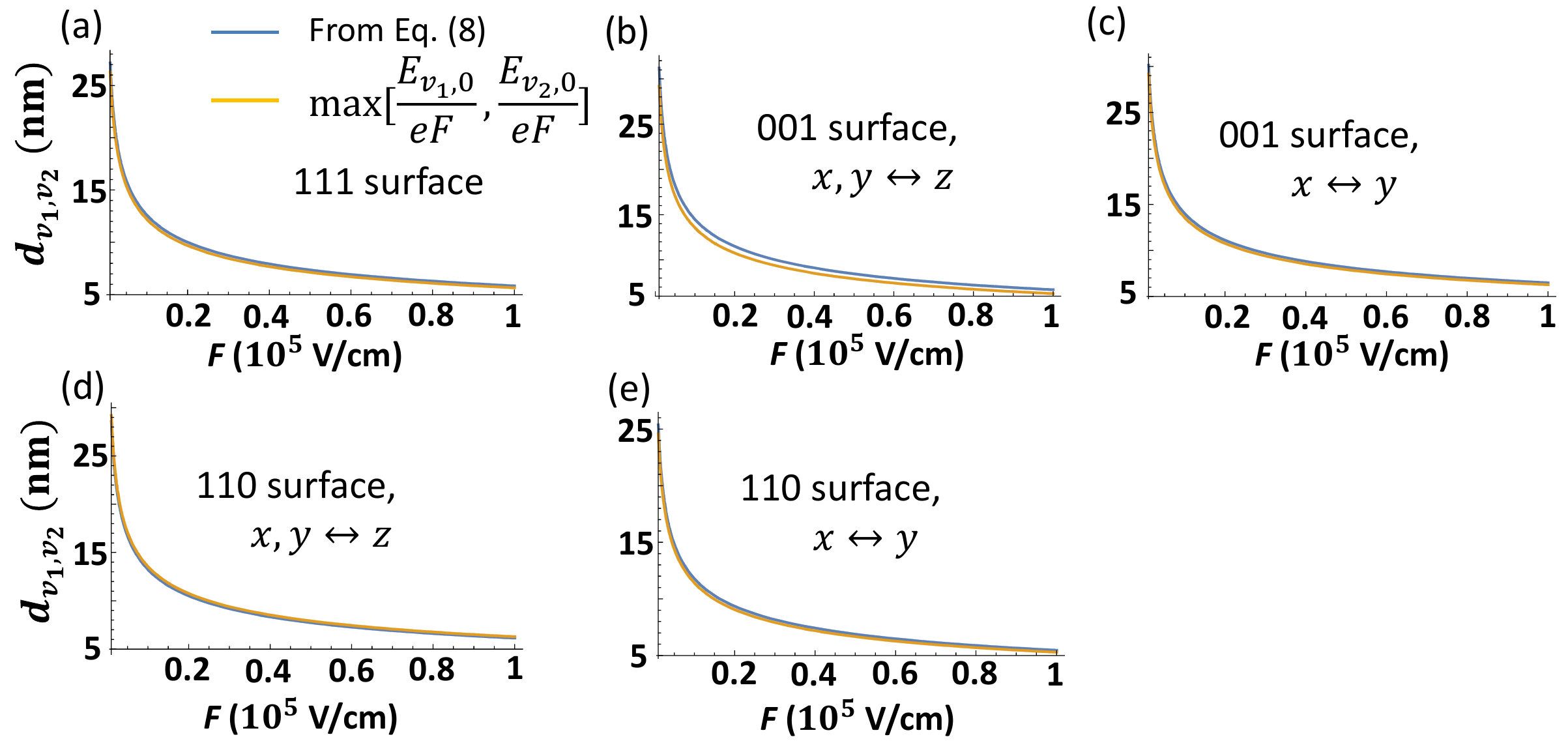}
\caption{ $d_{v_1,0;v_2,0}$ defined in Eq.~(\ref{eq:d_rev}) (the blue curves) as a function of effective electrical field $F$ for five representative cases of the triangular shape wells: $f$-process scattering (a) near the [111] surface, (b)  between the 4 and 2-valley groups near the [001] surface, (c) within the 4-valley groups near the [001] surface, (d)  between the 4 and 2-valley groups near the [110] surface, and (e) within the 4-valley groups near the [110] surface. In comparison, we also plot side by side max$[E_{v_1,0},E_{v_2,0}]/eF$ (the yellow curves).
}\label{fig:d_tr12}
\end{figure}

The second representative confinement we use for producing numerical results is a triangular well potential at the interface $V(z)=eFz$ for $z>0$ and $\infty$ for $z<0$ ($F$ is the electric field including built-in potential gradient). It corresponds approximately to the inversion (accumulation) layer of hole(electron)-doped Si MOSFET, when the 2DEG density is smaller than the saturated charge density of depletion layer per area, $N_{\rm depl}$ \cite{Ando_RMP82}.  We take the inversion layer as an example in Sec.~\ref{sec:detailed_results_for_2DEG}, while both types of 2DEG layers are treated in Sec.~\ref{sec:experimental_implications} with the variational approach. The envelope function in this model is analytically solved \cite{Stern_PRB72},
\begin{eqnarray}
\xi^{\rm tr}_{v,n}(z)\! &=& \!\alpha_{v,n} {\rm Ai}\!\left[\!  \left(\!\frac{2m_{z,v} e F}{\hbar^2}\!\right)^{\frac{1}{3}} \!\left(\!z-\frac{E_{v,n}}{eF}\!\right)\!\right]\! \theta(z),\quad\label{eq:eigenfunc_tri_well}
\\
E_{v,n} &\approx& \left(\frac{\hbar^2}{2m_{z,v}}\right)^{1/3} \left[ \frac{3}{2}\pi e F\left(n+\frac{3}{4} \right)\right]^{2/3} , \label{eq:eigenenergy_tri_well}
\end{eqnarray}
where  Ai denotes the Airy function, and $\alpha_{v,n}$ is the normalization factor [$y={\rm Ai}(x)$ satisfies the original Airy equation $y''-xy=0$].  $E_{v,n}$ are the asymptotic values for large $n$, but fall within 1\% of the exact value even for $n=0$. The Airy function depicts the oscillation of state envelopes within the classical turning point ($z_t=E_{v,n}/eF$) and the decay beyond it. The step function $\theta(z)$ [=1(0), for $z>(<)0$]  arises from the one-sided infinite barrier in the model.   For the triangular well, $d_{v_1,n_1;v_2,n_2}$ does not have a simple analytical form. In Fig.~\ref{fig:d_tr12}, we plot $d_{v_1,n_1;v_2,n_2} (F)$ in the typical range of electrical field $F$ ($10^3-10^5$ V/cm) numerically in the quantum limit, $n_1=n_2=0$,  for all representative surfaces and valley configurations. We find that for all these cases $d_{v_1,0;v_2,0}$ can be well approximated by
\begin{eqnarray}\label{eq:d_12_tr}
d^{\rm tr}_{v_1,0;v_2,0} \approx {\rm max}[\frac{E_{v_1,0}}{eF},\frac{E_{v_2,0}}{eF}].
\end{eqnarray}

The results in both Eqs.~(\ref{eq:d_12_sq}) and (\ref{eq:d_12_tr}), derived from the general definition of effective width in Eq.~(\ref{eq:d_rev}), can be physically interpreted as follows. First, the volume normalization of the initial and final states scales the scattering matrix element $U^{2d}$  inversely with $\sqrt{d_{v_1,n_1} d_{v_2,n_2}}$ where $d_{v,n}$ is the effective spread of the given subband state in the $z$ direction. Second, the relaxation rate scales with the number of impurities in the overlapping region of the two states, $\propto$ min$[d_{v_1,n_1},d_{v_2,n_2}]$.  Combining these two factors, the spin relaxation should be roughly proportional  to 1/max$[d_{v_1,n_1}, d_{v_2,n_2}]$ [which is effectively Eq.~(\ref{eq:d_rev})]. This is a generic prediction of our theory for impurity-induced 2DEG spin relaxation in Si, which could be directly tested experimentally. Finally, $d_{v,n}$ is basically $d$ for the square well and around the classical turning point $E_{v,n}/eF$ for the triangular well with slope $eF$.

With the core factors $U_{v_1,\mathbf{s}; v_2,-\mathbf{s}}$ and $d_{v_1,n_1; v_2, n_2}$ elaborated, in the following section we present our calculated spin relaxation results in all typical Si 2DEG orientations and stress configurations. By standard time-dependent perturbation theory, one integrates out the periodic time factors of the states resulting in the effective energy conservation in the large time limit, i.e., the Fermi golden rule \cite{Dirac_PRS27, Fermi_book50}, a standard application for relaxation rates, 
\begin{eqnarray}\label{eq:tau_s_2d}
\frac{1}{\tau^{2d}_s(\mathbf{s})} \!=\!
\frac{4\pi }{\hbar}\!
\bigg\langle\sum\limits_{v_2,n_2}  \!\int \!\frac{d^2k_2}{4\pi^2\!/\!S}N_{i} S  \!\int  \! \!dz |U^{2d}_{v_1,n_1; v_2,n_2}(z,\mathbf{s})|^2
\nonumber\\
\qquad\delta[\varepsilon_{v_1,n_1}(\mathbf{k}_1) \!-\! \varepsilon_{v_2,n_2}(\mathbf{k}_2)]
\bigg\rangle_{\mathbf{k}_1}
\end{eqnarray}
where $\mathbf{k}_1$ and $\mathbf{k}_2$  are the 2D wavevectors for initial and final states, $N_{i}$ is the impurity density per volume,  $\langle \mathcal{O}\rangle_{\mathbf{k}_1}\equiv \sum_{ \!v_1\!,n_1}\!\!\int \!d^2\! k_1 \mathcal{O}[\partial \!\mathcal{F}\!/\partial \varepsilon_{\!v_1\!,n_1\!}\!(\mathbf{k}_1\!)] \big/ \!\sum_{\!v_1\!,n_1}\!\!\int\! d^2 \!k_1 [\partial \mathcal{F}\!/\partial \varepsilon_{\!v_1\!,n_1\!}\!(\mathbf{k}_1\!)]$ denotes the shortcut for the normalized integration over $\mathbf{k}_1$   with $\mathcal{F}$ being the Fermi-Dirac distribution (see, e.g., Ref.~\cite{Yafet_SSP63}, p.~73). In our calculation, we neglect  the dependence of $U^{2d}_{v_1,n_1; v_2,n_2}$ on the small wavevector measured from its respective valley center ($\mathbf{k}_0$) \cite{Chalaev_arxiv16} which only renders a higher-order relative error [$\sim|\mathbf{k}-\mathbf{k}_0|/(2\pi/a)\ll 1$, $a$ being the Si lattice constant], and thus $U^{2d}$ depends only on the  valleys and subbands of the involved states.
Our leading-order in wavevector theory establishes the first quantitative analysis  for impurity-induced 2D spin relaxation beyond those arising from the DP mechanism.

\section{Numerical results for different 2DEG orientations and applied stress} \label{sec:detailed_results_for_2DEG}
\subsection{Without external stress}

In the limit of large well width, the number of occupied subbands for a given Fermi level is proportional to the width $d$, and $1/\tau^{2d}_s$ reduces to the 3D limit independent of $d$ \cite{Song_PRL14}, which we have explicitly verified numerically. In this section, we give concrete quantitative results for the opposite 2D limit of only one (``the quantum limit'') or few lowest subbands being populated, in the low-temperature limit of our interest.
As we stressed, the EMA is well applied in this limit for our problem where the interaction occurs within the impurity core regions. For more details on the general justification of applying EMA to tightly-confined quantum structures, the readers can refer to Ref.~\cite{Burt_JPCM92}. We also note that this theory treats on equal footing the ``weak-field'' and ``strong-field'' limits that arise from the study of structural SOC and oscillation of valley splitting \cite{Nestoklon_PRB08, Prada_NJP11}. For a square quantum well under strong electric field, the 2DEG may be modeled in a triangular confinement (or some specific variations) for our mechanism, as exemplified in Sec.~\ref{sec:theoretical formulation}.    We remark that since our spin-flip mechanism draws on $f$-process intervalley scattering,  it is easier to be seen for the (111) and (110)-oriented 2DEGs where non-opposite valleys coexist in the ground states, than for the (001) one. For the latter case, multi-subband occupation is required in order  for our leading-order spin relaxation to play a key role.

The situations with  potential confinement but no stress are studied first.  $\tau_s^{2d}$ as a function of 2DEG electron density $N_{2d}$ for a given well potential $V(z)$, as well as $\tau_s^{2d}$ versus $V(z)$ for a given $N_{2d}$, are computed. Depending on  $N_{2d}$ and the subband splitting (or the corresponding well width), different subbands may be populated.  Unlike the 3D bulk case, in 2DEG the electron density and the Fermi level are decoupled from the impurity density $N_i$ as the former can be controlled by the gate voltage. The general relation between $N_{2d}$  and the Fermi level $\varepsilon_F$ reads,
\begin{eqnarray}
N_{2d}\! = \frac{ 1}{2\pi \hbar^2}\!\sum_v \!\sum_n \!\sqrt{m_{1,v} m_{2,v}} (\varepsilon_F \!-\! E_{v,n}) \theta\!(\varepsilon_F \!-\! E_{v,n}),
\end{eqnarray}
where $m_{1,v}$ and $m_{2,v}$ are the in-plane effective masses in the $v$th valley (the effective mass is anisotropic in Si due to the ellipsoidal forms of the bulk conduction band minima), and $\theta(x)=0$ or 1 for $x<0$ or $x>0$.

We start with 2DEG of  the (111) well orientation. $m_z=3m_t m_l/(m_t+2 m_l)$ is the same in every valley, so the six energy ``ladders'' of subbands remain degenerate among different valleys (neglecting any small valley splitting correction beyond the effective mass approximation).   The spin relaxation rate, Eq.~(\ref{eq:tau_s_2d}), then becomes,
\begin{eqnarray}\label{eq:tau_s_2d_111}
\frac{1}{\tau^{(111)}_s(\mathbf{s})} =
\frac{\pi^2 a^6_B \Delta^2_{\rm so} N_i }{\hbar^3 } \mathcal{G}^{(111)},
\end{eqnarray}
with the orientation-specific factor assuming the low-temperature limit,
\begin{eqnarray}\label{eq:factor_111}
\mathcal{G}^{(111)} \! = \!
\frac{\sum\limits_{n_1,n_2} \frac{4(1+6\eta^2)\sqrt{ m_1 m_2} }{9d_{n_1,n_2}}  \theta(\varepsilon_F-E_{n_1}) \theta(\varepsilon_F-E_{n_2})}
{3\sum_n \theta(\varepsilon_F-E_{n})},
\end{eqnarray}
where $m_1=m_t$ and $m_2=(m_t+2m_l)/3$.  As in the 3D case, $\tau^{(111)}_s(\mathbf{s})$ is isotropic in spin orientation.

\begin{figure}[!htbp]
\centering
\includegraphics[width=8.5 cm]{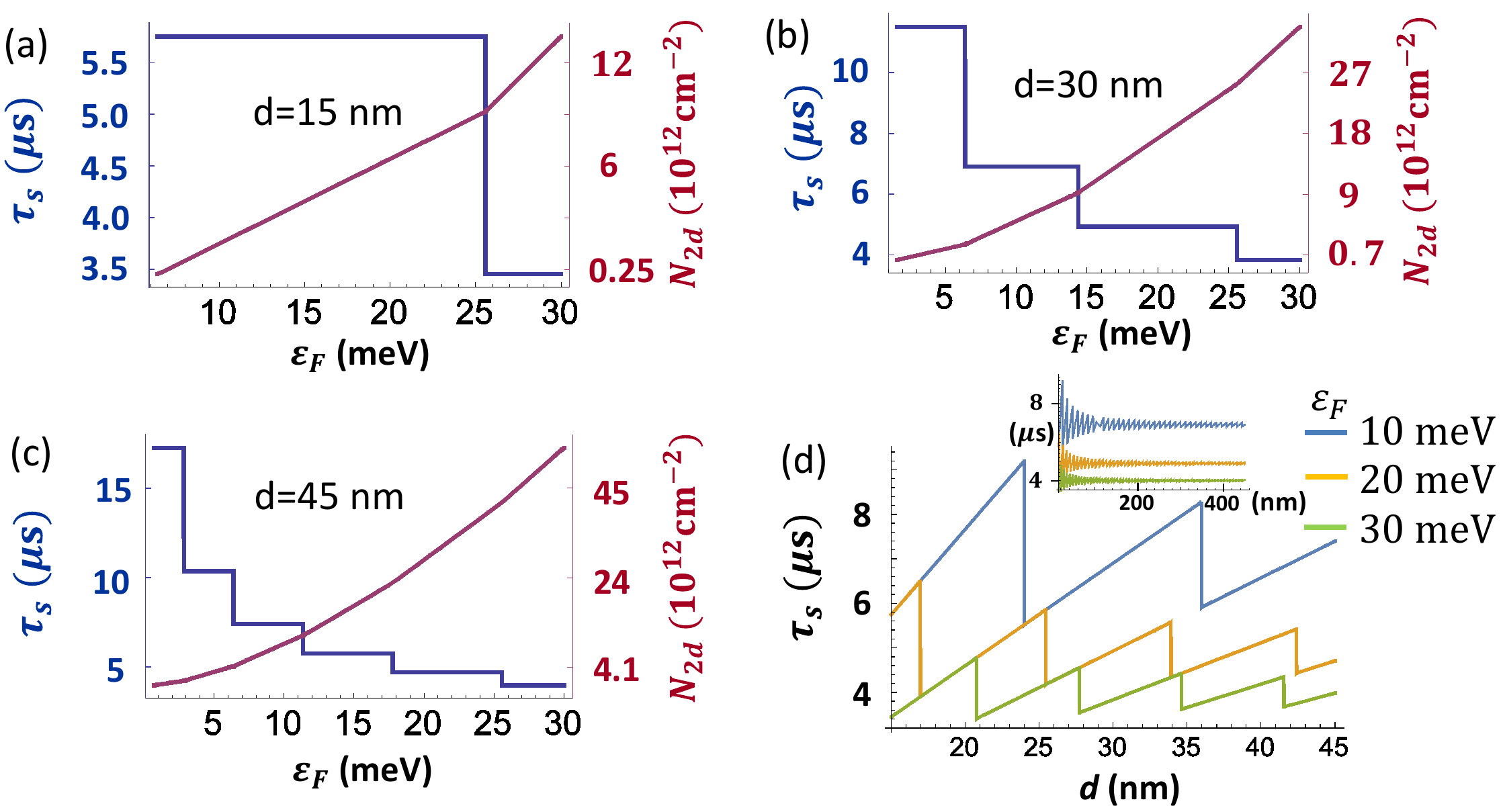}
\caption{$\tau_s$ in the (111) square well. We plot $\tau_s$ as a function of $\varepsilon_F$ and the corresponding $N_{2d}$ for three  representative well widths $d$, (a) 15 nm, (b) 30 nm, and (c) 45 nm, and  $\tau_s$ as a function of $d$ for three fixed $\varepsilon_F=10,20$ or 30 meV in (d). Inset of (d) shows the large $d$ behavior  approaching the bulk limit. Note that the energy of the lowest subband bottom is $E_0=\pi^2\hbar^2/2m_z d^2$ relative to the zero reference energy. $\Delta_{\rm so}=0.1$ meV and $N_i=10^{16}$ cm$^{-3}$ are chosen here and for all the following figures.
}\label{fig:tau_s_111surf_sq}
\end{figure}

We quantify the spin relaxation time $\tau_s$ for the two basic well types introduced in Sec.~\ref{sec:theoretical formulation}, (1) the infinite square well and (2) the triangular well.
For a square well with width $d$, $E_n$ and $d_{n_1,n_2}$ follow Eqs.~(\ref{eq:E_n_sq}) and (\ref{eq:d_12_sq}), and Eq.~(\ref{eq:factor_111}) reduces to
\begin{eqnarray}\label{eq:factor_111_sq}
\mathcal{G}^{(111)}_{\rm sq}\! \!   =\!
\frac{4(1\!+ \!6\eta^2)\! \sqrt{ m_1 m_2} }{27d}  \! \left[\! \mathcal{N}\! \left(\!\! \sqrt{\frac{\varepsilon_F}{E_0}}\right)\!+ \!\frac{1}{2}\right] \theta(\varepsilon_F \!-\! E_0),\;\;
\end{eqnarray}
where $\mathcal{N}(x)$ returns the integer part of $x$.
We plot $\tau_s$ as a function of $\varepsilon_F$ and relate it to the corresponding $N_{2d}$ in Fig.~\ref{fig:tau_s_111surf_sq} for three different representative well widths, (a) 15 nm, (b) 30 nm, and (c) 45 nm. We also plot $\tau_s$ as a function of $d$ for three fixed  $\varepsilon_F=10,20$ or 30 meV in Fig.~\ref{fig:tau_s_111surf_sq}(d). Since one can simply scale $\tau_s$ with $\Delta_{\rm so}^{-2}$ and $N_i^{-1}$ as shown in Eq.~(\ref{eq:tau_s_2d_111}), we choose typical parameters $\Delta_{\rm so}=0.1$ meV  and doping concentration $N_i=10^{16}$ cm$^{-3}$. The clear kinks in $N_{2d}$ versus $\varepsilon_F$ and the jumps in $\tau_s$ versus $\varepsilon_F$ or $d$ reflect the onset of (de)populating more subbands. $\tau_s$ decreases as $\varepsilon_F$ or $N_{2d}$ increases for a fixed $d$, since more subbands are available for states at the Fermi level to be scattered into. As $d$ increases towards the bulk limit, denser subbands gradually evolve towards the density of state for  3D bulk at a given $\varepsilon_F$ [see inset of Fig.~\ref{fig:tau_s_111surf_sq}(d)].   On the other hand, at small $d$, $\tau_s$ decreases continuously with decreasing $d$ within the window of  a fixed number of occupied subbands, $n_{occ}$. This is a general trend dominated by the volume normalization of the involved state captured in Eq.~(\ref{eq:d_rev}). When $n_{occ}$ reduces by one, the number of available final states decreases and $\tau_s$ increases again.

\begin{figure}[!htbp]
\centering
\includegraphics[width=8.cm]{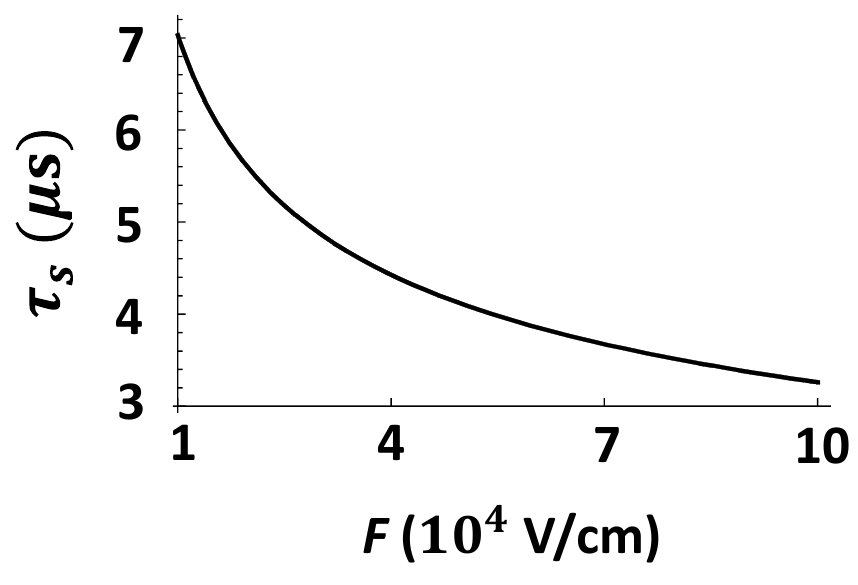}
\caption{$\tau_s$ in the triangular well $V(z)=eFz$ to the (111) surface for $F\leq 10^5$ V/cm. This is at the quantum limit where only the lowest subbands are occupied. In this limit, $\tau_s$ is independent of $\varepsilon_F$ and  $N_{2d}$. We choose the realistic acceptor density in the depletion as well as in the inversion layer, $N_i=N_A=10^{16}$ cm$^{-3}$.
}\label{fig:tau_s_111surf_tri}
\end{figure}

For the triangular well $V(z)=eFz$, with its slope controllable by the gate voltage in an inversion layer, the solution of $\xi_n(z)$ becomes the Airy functions given in Eq.~(\ref{eq:eigenfunc_tri_well}), and $d^{\rm tr}_{n_1,n_2}$ for $n_1=n_2=0$ can be estimated by Eq.~(\ref{eq:d_12_tr}).  The general changes from the square well are (1) $d_{n_1;n_2}$ in Eq.~(\ref{eq:tau_s_2d_111}) becoming $n_{1,2}$ dependent, and (2) the different dependence of $E_n$ on $n$.  However, we emphasize that the triangular model is more valid as $N_{2d}/N_{\rm depl}$ decreases (and $N_{2d}<N_{\rm depl}$) \cite{Ando_RMP82}. A  realistic acceptor density we choose is $N_A=10^{16}$ cm$^{-3}$. As a result, this usually corresponds to the situation where only the lowest subbands in the inversion layer are occupied. Therefore the numerical results in Fig.~\ref{fig:tau_s_111surf_tri} is given in this practical energy window. In this case, $\tau_s$ becomes independent of $\varepsilon_F$ or $N_{2d}$ due to the constant 2D density of states per subband, and we only need to show the dependence of $\tau_s$ on $F$. For higher $N_{2d}$, $V(z)$ is not independent of but largely determined by $N_{2d}$.

Next we study the (001) well orientation, where additionally we have the relative shift between different subbands belonging to the 2-valley group ($\pm z$ valleys in the 3D limit) and 4-valley group ($\pm x$ and $\pm y$ valleys in the 3D limit). The subband edges in these two groups are determined by the different $m_z$'s, $m_z=m_l (m_t)$ for the 2(4)-valley group. This symmetry breaking between the 6 otherwise equivalent ladders of subbands results in spin-orientation dependence, absent in the (111) well case. In the quantum limit, only the 2-valley group is occupied and the spin relaxation due to impurities vanishes in the leading order.    The general spin relaxation rate follows Eq.~(\ref{eq:tau_s_2d_111}) with $\mathcal{G}^{(111)}$ replaced by
\begin{eqnarray}\label{eq:factor_001}
\mathcal{G}^{(001)}(\mathbf{s}) \!\!&=&\!\!
\sum\limits_{n_1, n_2} \!\! \theta(\varepsilon\!_F\!-\!E_{x,n_2})\!\bigg\{\!\frac{\mathcal{S}(s_z) \sqrt{m_t m_l} \theta(\varepsilon\!_F\!\!-\!E_{x,n_1}) }{d_{x,{n_1};y,{n_2}}}
\quad\nonumber\\
&&+
\frac{ [ \frac{4}{9}(1+6\eta^2)-\mathcal{S}(s_z)] m_t  \theta(\varepsilon_F\!-\!E_{z,n_1})}{d_{x,{n_1};z,{n_2}}} \bigg\}
\nonumber\\
&&\!\!\!\!\!\!\!\!\bigg/\!
 \sum_{n}\left[\sqrt{\frac{m_t}{m_l}}   \theta(\varepsilon_F\!-\!E_{z,n}) + 2 \theta(\varepsilon_F\!-\!E_{x,n})\right] ,
\end{eqnarray}
after utilizing the spin-orientation form factors in Eqs.~(\ref{eq:U_xy})-(\ref{eq:U_zy}). We have used the anisotropic in-plane effective masses:  $m_1=m_2=m_t$ for the 2-valley group, and $m_1=m_t, m_2=m_l$ for the 4-valley group. For square well with width $d$, we have
\begin{eqnarray}\label{eq:factor_001_sq}
\mathcal{G}^{(001)}_{\rm sq}(\mathbf{s})\!\!&=&\!\!
\frac{\theta(\varepsilon_F\!-\!E_{x,0})}{d}  \left[\mathcal{N}\!\left(\!\sqrt{\frac{\varepsilon_F}{E_{x,0}}}\right)+\frac{1}{2}\right]
\nonumber\\
&&
\bigg \{  \mathcal{S}(s_z) \sqrt{m_t m_l}\; \mathcal{N}\!\left(\!\sqrt{\frac{\varepsilon_F}{E_{x,0}}}\right)
\nonumber\\
 &&+
  [ \frac{4}{9}(1+6\eta^2)-\mathcal{S}(s_z)] m_t \;\mathcal{N}\left(\!\sqrt{\frac{\varepsilon_F}{E_{z,0}}}\!\right)\!\!
\bigg\}
\nonumber\\
&& \bigg/\!\left[\sqrt{\frac{m_t}{m_l}}\;  \mathcal{N}\!\left(\!\sqrt{\frac{\varepsilon_F}{E_{z,0}}}\right) + 2   \;\mathcal{N}\!\left(\!\sqrt{\frac{\varepsilon_F}{E_{x,0}}}\right) \!\right]. \qquad
\end{eqnarray}

\begin{figure}[!htbp]
\centering
\includegraphics[width=8.5cm]{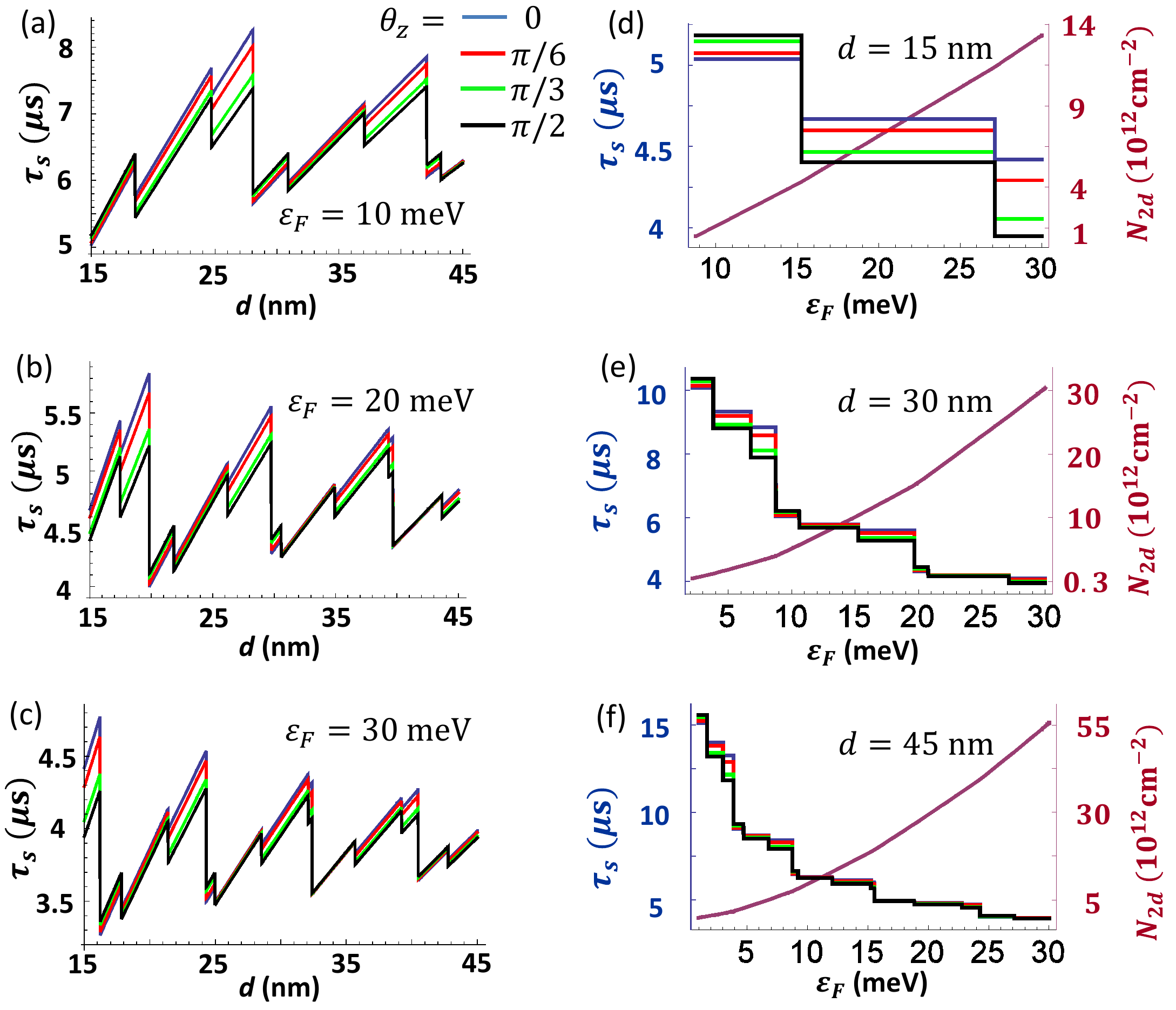}
\caption{$\tau_s(\mathbf{s})$ in the (001) square well, anisotropic in spin orientation $\mathbf{s}$. $\tau_s$ depends on $\mathbf{s}$'  polar angle, $\theta_z$, with respective to the well normal, $z$.  In (a)-(c), we vary the well width $d$ from 15 to 45 nm for three Fermi levels $\varepsilon_F=$(a) 10, (b) 20 and (c) 30 meV.  In (d)-(f), we plot $\tau_s(\mathbf{s})$ as a function of $\varepsilon_F$, $E_{x,0}<\varepsilon_F<30$ meV (and also a function of $N_{2d}$ by relating $\varepsilon_F$ with $N_{2d}$), for three different well widths, $d=$ (d) 15, (e) 30 and  (f) 45 nm.  In each subplot we exemplify four spin polar angles $\theta_z=0,\pi/6,\pi/3$ and $\pi/2$.
}\label{fig:tau_s_001surf_sq}
\end{figure}

Plots of $\tau_s(\mathbf{s})$ with $\mathcal{G}^{(001)}_{\rm sq}(\mathbf{s})$ in Eq.~(\ref{eq:factor_001_sq}) as functions of $d$ and $\varepsilon_F$ (and the corresponding $N_{2d}$) are given in Fig.~\ref{fig:tau_s_001surf_sq}. An apparent new feature is the generally smaller steps in comparison with the (111) well results,  resulting from the consecutive fillings of the 2-valley subbands which have smaller interband splitting. As mentioned, the important consequence of the inequivalency between the 2-valley and 4-valley groups is the spin-orientation dependence of the spin lifetime. The anisotropy is the strongest when the occupied states in the two groups differ the most, which happens right before one more 4-valley subband begins to be filled. We need to note that merely $n_{occ,z}> n_{occ,x}$ is not enough to guarantee spin anisotropy, but it has to be  $n_{occ,z}/n_{occ,x}> \sqrt{m_{1,x} m_{2,x}/ m_{1,z} m_{2,z}}=\sqrt{m_l/m_t}$ based on Eq.~(\ref{eq:factor_001_sq}).  This is most appreciable preceding the filing of the second subband in the 4-valley group, $n_{occ,z}/n_{occ,x}=4:1$. In the large $\varepsilon_F$ (or large $d$) limit, on the other hand, $n_{occ,z}/n_{occ,x}\rightarrow \sqrt{m_{z,z}/ m_{z,x}}= \sqrt{m_l/m_t}\approx 2.27$. We see that this ratio exactly cancels out the effective mass difference in the 2D ($x$-$y$) plane, owing to the fact that $m_1 m_2 m_z\equiv m_l m_t^2$ is orientation-independent for a given ellipsoid.

\begin{figure}[!htbp]
\centering
\includegraphics[width=8.5cm]{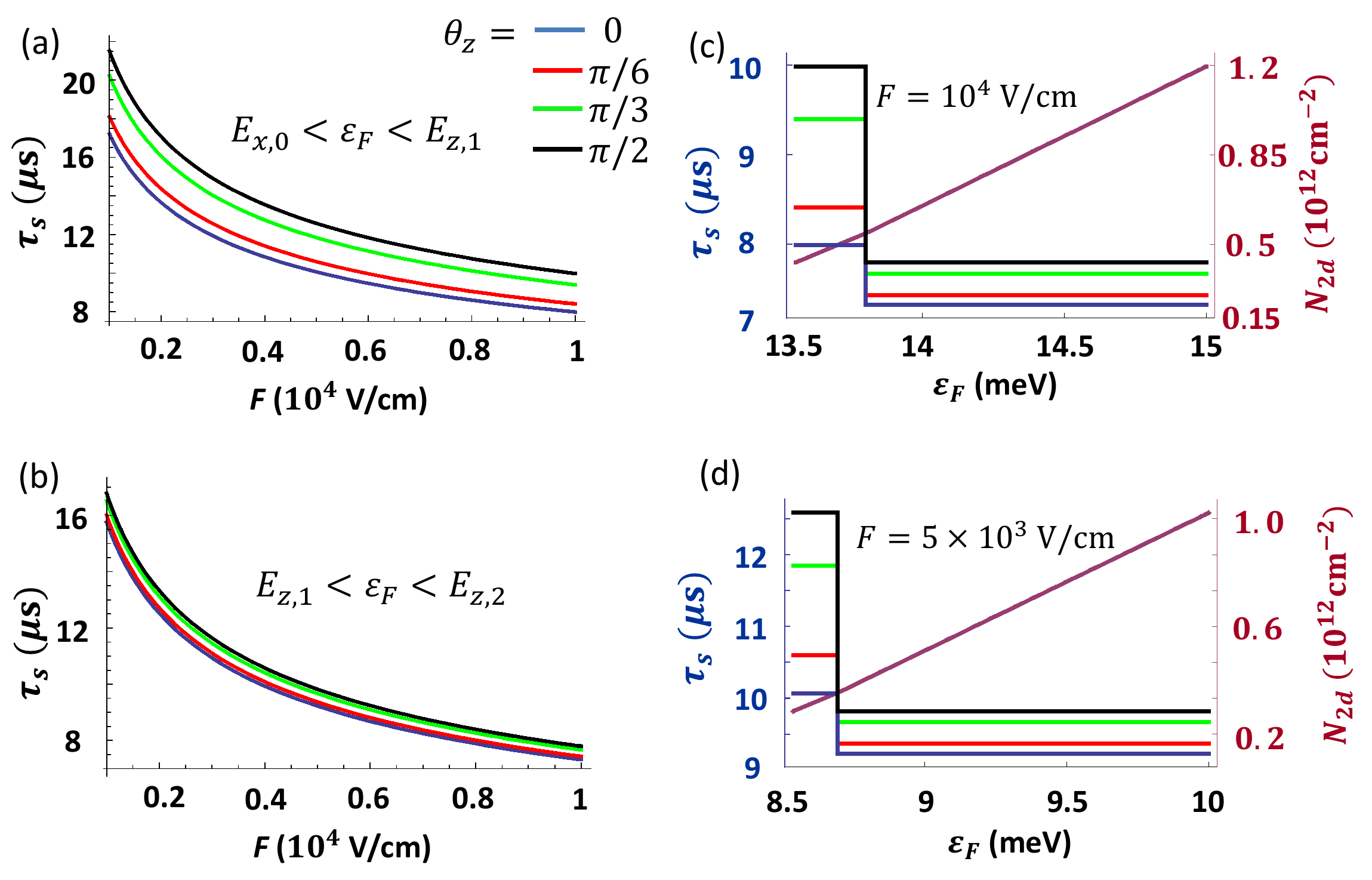}
\caption{$\tau_s(\mathbf{s})$ in the triangular well $V(z)=eFz$ to the (001) surface. In (a) and (b), $\tau_s(\mathbf{s})$ is plotted as a function of electric field $F<10^4$ V/cm for the Fermi level $\varepsilon_F$ (a) just above the onset of $f$-process scattering ($E_{x,0}<\varepsilon_F< E_{z,1}$), and (b) in the next energy window ($E_{z,1}<\varepsilon_F< E_{z,2}$), where $\tau_s(\mathbf{s})$ is independent of $\varepsilon_F$. In (c) and (d), $\tau_s(\mathbf{s})$ is plotted as a function of $\varepsilon_F$ and $N_{2d}$ for a range $E_{x,0}<\varepsilon_F< E_{z,2}$, at electric field $F=$ (c) $10^4$ and (d) $5\times 10^3$ V/cm. At $F=10^4$ V/cm, the onset of $f$-process scattering already requires $N_{2d}\approx 5\times 10^{11}$ cm$^{-2}$, a comparable value to the typical depletion layer impurity density $N_{\rm depl}=\sqrt{2 E_G \kappa N_A/e^2}$ \cite{Ando_RMP82} where $E_G$ and $\kappa$ are the Si band gap and permittivity ($N_{\rm depl}\sim 4\times 10^{11}$ cm$^{-2}$ at $N_A=10^{16}$ cm$^{-3}$).
}\label{fig:tau_s_001surf_tri}
\end{figure}

For the triangular well, Eqs.~(\ref{eq:d_rev}), (\ref{eq:eigenfunc_tri_well}) and (\ref{eq:eigenenergy_tri_well}) are substituted into Eq.~(\ref{eq:factor_001}), and $\tau_s(\mathbf{s})$ is shown in Fig.~\ref{fig:tau_s_001surf_tri}. After the onset of $f$-process scattering ($\varepsilon_F>E_{x,0}$) follows the second subband of the 2-valley group, as shown clearly in Fig.~\ref{fig:tau_s_001surf_tri}(c) and (d). $(E_{z,1}- E_{x,0})/E_{x,0}$ is a small fixed ratio for any electric field $F$ according to Eq.~(\ref{eq:eigenenergy_tri_well}). Since we work in the regime where the triangular well model is valid and multiple subbands are occupied,  $\varepsilon_F$ and $F$ should not be too large. As a result, we focus on two energy intervals, $E_{x,0}<\varepsilon_F< E_{z,1}$ and $E_{z,1}<\varepsilon_F< E_{z,2}$ below $F=10^4$ V/cm (in practice, our mechanism works well under higher electrical field, as long as the Fermi level can reach the $x$ and $y$ valleys and the triangle well approximation is relaxed).

For scattering involving subband $n=1$, we obtain $d_{z,1;x,0}\approx 1.15 E_{z,1}/eF$ using Eqs.~(\ref{eq:d_rev}) and (\ref{eq:eigenfunc_tri_well}). An interesting behavior is the large anisotropy ($\sim 20\%$) of $\tau_s(\mathbf{s})$ right at the onset of the $f$-process scattering [Fig.~\ref{fig:tau_s_001surf_tri}(a),(c) and (d)], which drops ($\sim 5\%$) at the second energy window [cf. Fig.~\ref{fig:tau_s_001surf_tri}(b)]. In the first energy window, $E_{x,0}<\varepsilon_F< E_{z,1}$,  the number of occupied subbands in each valley is one for both 2-valley and 4-valley groups. The large anisotropy here is the sole consequence of effective mass anisotropy in the 2D plane and is opposite in sign to that in the square well before filling $E_{x,1}$.

The change of anisotropy may be sharply \textit{tuned by the gate voltage} in Si inversion layer. Since $E_{z,1}- E_{x,0}$ and $\varepsilon_F-E_{x,0}$ depend on the electric field $F$, by just tuning $F$ the chemical potential (i.e., Fermi energy) can cross $E_{z,1}$ and therefore induce a switch between Fig.~\ref{fig:tau_s_001surf_tri}(a) and (b). An even more important application of our results may be the sharply \textit{gate-voltage modulated}  spin lifetime in Si inversion layer following a similar reasoning: the crossover of chemical potential to $E_{x,0}$, which is the threshold of finite leading-order spin relaxation, could be achieved solely by the top gate voltage. This is the on-chip real-time electrical switch for a substantial change of the spin lifetime.  We emphasize that both gate-voltage modulations envisioned above should be very robust in general Si inversion layers, not relying on the triangular well approximation we used in producing Fig.~\ref{fig:tau_s_001surf_tri}.

Last, we study the case of the (110) well. The difference of the (110) well from the (001) one lies in the effective masses, which result in a quantitative difference in the valley splitting, the subband splitting and the density of states, all playing roles in determining $\tau_s$. For the 2-valley group, $m_z=m_t$, $m_1=m_t$ and $m_2=m_l$; for the 4-valley group, $m_z= 2m_t m_l/(m_t+m_l)$, $m_1=m_t$ and $m_2=(m_t+m_l)/2$. The most important distinction is in the quantum limit where the lowest subbands in the (110) case  are from the 4-valley group, rather than from the 2-valley group in the (001) case. Consequently, the spin lifetime due to the leading-order spin relaxation is finite even for the lowest electron density in the (110) well.

\begin{figure}[!htbp]
\centering
\includegraphics[width=8.5cm]{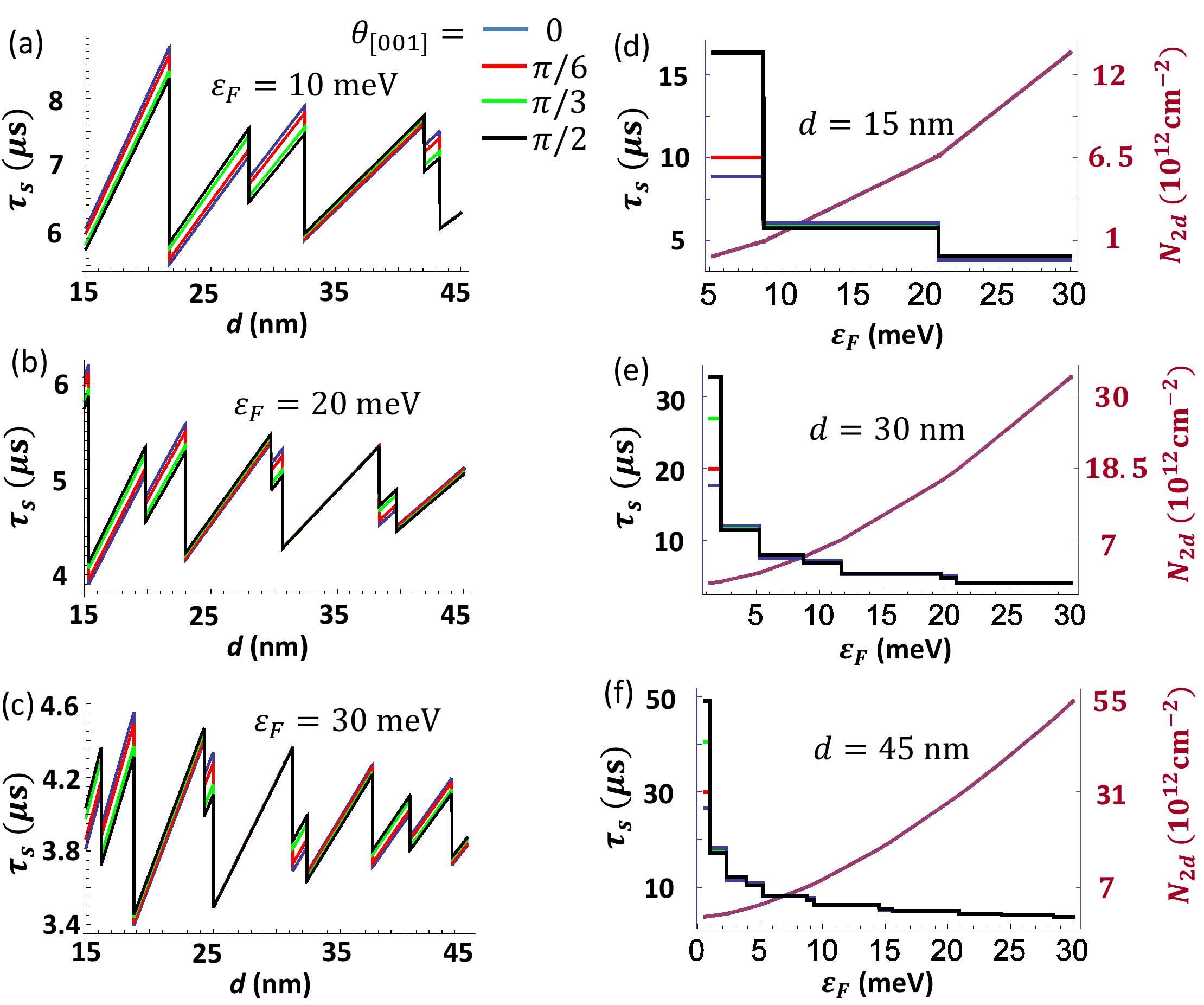}
\caption{$\tau_s(\mathbf{s})$ in the (110) square well, anisotropic in spin orientation $\mathbf{s}$. $\tau_s$ depends on $\mathbf{s}$' polar angle, $\theta_{[001]}$, with respective to the [001] crystallographic direction.  In (a)-(c), we vary the well width $d$ from 15 to 45 nm for three Fermi levels $\varepsilon_F=$ (a) 10, (b) 20 and (c) 30 meV. In (d)-(f), we plot $\tau_s(\mathbf{s})$ as a function of $\varepsilon_F<30$ meV and  $N_{2d}$, for three different well widths, $d=$ (d) 15, (e) 30 and  (f) 45 nm.
}\label{fig:tau_s_110surf_sq}
\end{figure}

\begin{figure}[!htbp]
\centering
\includegraphics[width=8.5cm]{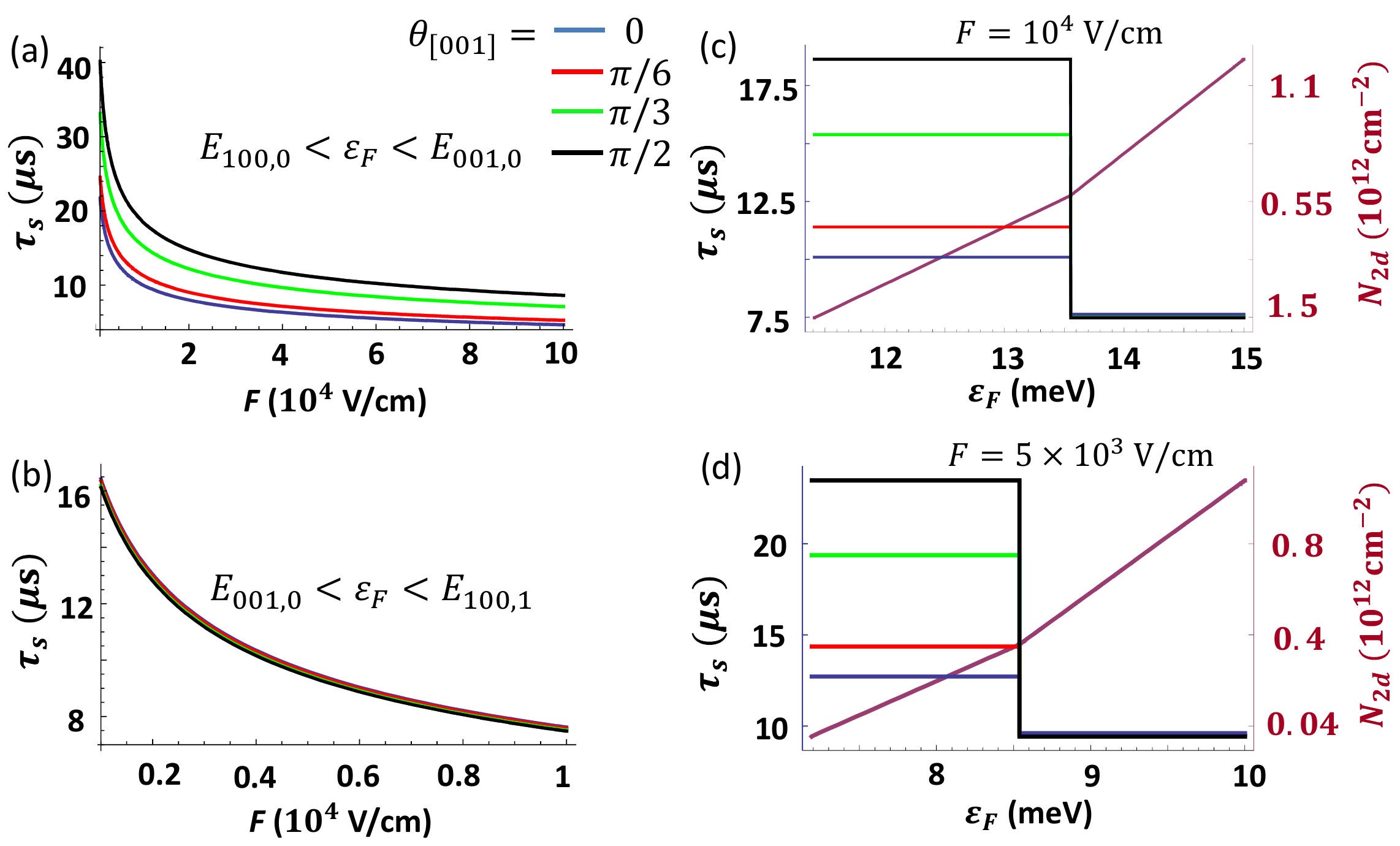}
\caption{$\tau_s(\mathbf{s})$ in the triangular well $V(z)=eFz$ to the (110) surface. In (a) and (b), $\tau_s$ is plotted as a function of the electric field $F$ for the Fermi level $\varepsilon_F$ (a) in the lowest energy interval ($E_{100,0}<\varepsilon_F< E_{001,0}$) with $F<10^5$ V/cm , and (b) in the next energy window ($E_{001,0}<\varepsilon_F< E_{100,1}$) with $F<10^4$ V/cm , where $\tau_s(\mathbf{s})$ is independent of $\varepsilon_F$. We work in a smaller field in (b) for the same reason in Fig.~\ref{fig:tau_s_001surf_tri} (a) and (b).  In (c) and (d), $\tau_s$ is plotted as a function of $\varepsilon_F$ and  $N_{2d}$ for a range $\varepsilon_F< E_{100,1}$, at electric field $F=$ (c) $10^4$ and (d) $5\times 10^3$ V/cm.
}\label{fig:tau_s_110surf_tri}
\end{figure}

We show the key results for the (110) square well in Fig.~\ref{fig:tau_s_110surf_sq} and the triangular well in Fig.~\ref{fig:tau_s_110surf_tri}. We use similar parameters (Fermi level $\varepsilon_F$, well width $d$, electric field $F$) as those in the (001) case, so that we can focus on the differences between (110) and (001) wells. First,  $\tau_s$ is finite in the (110) 2DEG even for the lowest $\varepsilon_F$ in Fig.~\ref{fig:tau_s_110surf_sq} (d)-(f) and Fig.~\ref{fig:tau_s_110surf_tri} (c) and (d), as mentioned above. Due to the smaller number of available $f$-process paths (two as opposed four for each state), though,  $\tau_s$ is larger in the (110) case than the longest finite $\tau_s$ in the (001) case. This is a clearly verifiable sharp prediction of our theory. Note that here the plane normal ($z$) is along the crystallographic direction [110] (not [001]). To avoid ambiguity, we use scripts 001 or 100 rather than $z$ or $x$ to denote directions.

The most significant feature in this quantum limit for (110) wells is the nearly 50\% variation of $\tau_s (\mathbf{s})$ on spin orientation (more specifically, on $\mathbf{s}$' projection along [001] crystallographic direction). This is the extreme case of only one type of $f$-process scattering [Eq.~(\ref{eq:U_xy})] with zero weight from the other two. This feature is clearly seen in the left side of Fig.~\ref{fig:tau_s_110surf_sq}(d)-(f), Fig.~\ref{fig:tau_s_110surf_tri} (c) and (d), and the entire range in Fig.~\ref{fig:tau_s_110surf_tri}(a). Therefore, the idea of gate-tuned anisotropic $\tau_s (\mathbf{s})$, discussed in the context of (001) wells, is even more prominent in (110) wells. Note that this dependence on the polar angle around [001] direction, $\theta_{[001]}$, is opposite in sign to that of the (001) square well when the   anisotropy is the strongest and the same as that in the (001) triangular well. The anisotropy and the orientation dependence of 2D Si spin relaxation arising in the impurity-induced spin-flip is an important prediction of our theory.

\subsection{With external stress}

When external stress is applied, different ladders of subbands may undergo relative shift with each other \cite{DasSarma_PRB79}. On the other hand, no leading-order effect comes from the slight variation of interband splitting within each ladder, as the effective masses are fixed under stress to the leading order  \cite{DasSarma_PRB79}. We study $\tau_s$ individually for all three well orientations under uniaxial stress as described below.

\textit{(111) well}. Out-of-plane [111] stress does not induce additional symmetry breaking or shift subband ladders, while the in-plane stress in $[1\bar{1}0]$ ($\bar{1}\equiv -1$) or $[1-\sqrt{3},1+\sqrt{3},-2]$ direction does. These latter two stress directions are in the plane of the 2D well  and realizable experimentally \cite{Dorda_PRB78}. The $[1\bar{1}0]$ (or $[11\bar{2}]$)  stress results in a relative shift $\Delta_V= E_{001}- E_{100}$ between the 2-valley and 4-valley groups, and the $[1-\sqrt{3},1+\sqrt{3},-2]$ one separates the valleys into three groups such that, $\Delta_V=E_{010}-E_{001} = E_{001}-E_{100}$.

Under the $[1\bar{1}0]$ stress, the form factor $\mathcal{G}$ changes from Eq.~(\ref{eq:factor_111}) for the unstrained (111) well to,
\begin{eqnarray}\label{eq:tau_s_111surf_1-10}
\mathcal{G}^{(111)}_{1\bar{1}0}(\mathbf{s}) \!\!&=&\!\!
\sqrt{\frac{m_t(m_t+2 m_l)}{3}}
\\
&&\!\!\!
\sum\limits_{n_1, n_2} \!\frac{ \theta(\varepsilon_F-E_{100,n_2})}{d_{{n_1};{n_2}}} \bigg\{\mathcal{S}(s_z)   \theta(\varepsilon_F\!\!-E_{100,n_1})
\nonumber\\
&&\!\!\!\!+\!
  \left[ \frac{4}{9}(1+6\eta^2)\!-\!\mathcal{S}(s_z)\right]\theta(\varepsilon_F\!-\!E_{100,n_1}\!-\!\Delta_V)\!\bigg\}
\nonumber\\
&&
\!\!\!\bigg/\!\sum_{n}\left[   \theta(\varepsilon_F\!-\!E_{100,n}\!-\!\Delta_V)\! +\! 2 \theta(\varepsilon_F\!-\! E_{100,n}) \right] .\nonumber
\end{eqnarray}
Focusing on the near quantum limit with only $n=0$ subbands occupied, $d_{0,0}=2d/3$ for the square well and about $( 9\pi \hbar)^{2/3}/[4(2m_{111} eF)^{1/3}]$ for the triangular well, by Eqs.~(\ref{eq:d_12_sq}) and (\ref{eq:d_12_tr}) respectively.  Equation~(\ref{eq:tau_s_111surf_1-10}) has only a few discrete outcomes for a given well width or electric field: (1) when $\varepsilon_F>\{E_{100,0}, E_{100,0}+\Delta_V
\}$,  lines 2-4 of Eq.~(\ref{eq:tau_s_111surf_1-10}) reduce to $4(1+6 \eta^2)/27 d_{0;0}$ and $\tau_s$ recovers the no-strain result (Figs.~\ref{fig:tau_s_111surf_sq} and \ref{fig:tau_s_111surf_tri}); (2) when  $E_{100,0}<\varepsilon_F < \Delta_V+E_{100,0}$, the same factor decreases to $[1-s_z^2+3\eta^2(1+s_z^2)]/9 d_{0;0}$ with a strong $s_z$  dependence; (3) when  $\Delta_V+E_{100,0}<\varepsilon_F<E_{100,0} $, $1/\tau_s=0$.

Under the $[1-\sqrt{3},1+\sqrt{3},-2](\equiv \bm\gamma)$ stress, three groups of $f$-process scattering vary independently. Utilizing Eqs.~(\ref{eq:U_xy})-(\ref{eq:U_zy}), the form factor $\mathcal{G}^{(111)}_{\bm\gamma}$ reads,
\begin{eqnarray}\label{eq:tau_s_111surf_-13-2}
\mathcal{G}^{(111)}_{\bm\gamma}(\mathbf{s}) \!\!&=&\!\!
\sqrt{\frac{m_t(m_t+2 m_l)}{3}}
\nonumber\\
&&\!\!\!\frac{
\sum\limits_{n\!_1, n\!_2,i} \!\!\frac{ \mathcal{S}(s_i)}{d_{{n_1};{n_2}}}  \theta(\varepsilon_F\!\!-E_{i+1,n_1}) \theta(\varepsilon_F\!-\!E_{i+2,n_2})
}
{\sum_{n,i} \theta(\varepsilon_F\!-\!E_{i,n}) }.\qquad
\end{eqnarray}
where $i$ denotes the 3 cyclic directions $100,010$ and $001$, $E_{010,n}=E_{001,n}+\Delta_V$ and $E_{100,n}=E_{001,n}-\Delta_V$.  Although three $f$-process groups depend on spin projection along different directions, shown  in Eqs.~(\ref{eq:U_xy})-(\ref{eq:U_zy}), $\tau_s(\mathbf{s})$ can be associated with a fixed projection direction in each energy window of $\varepsilon_F$, thanks to the constant density of state per subband. Focusing on the $n=0$ limit, the second line in Eq.~(\ref{eq:tau_s_111surf_-13-2}) has several possible outcomes: (1) $4(1+6 \eta^2)/27 d_{0;0}$,  when $\varepsilon_F>\{E_{100,0}, E_{010,0}, E_{001,0}\}$, recovering the no-strain result; (2) $\mathcal{S}(s_{100})/2 d_{0;0}$, when $\{E_{010,0}, E_{001,0}\}<\varepsilon_F< E_{100,0}$; (3) $\mathcal{S}(s_{010})/2 d_{0;0}$, when $\{E_{100,0}, E_{001,0}\}<\varepsilon_F< E_{010,0}$; and finally (4) 0, when $ E_{100,0}<\varepsilon_F<\{E_{010,0}, E_{001,0}\}$ or $E_{010,0}<\varepsilon_F<\{E_{100,0}, E_{001,0}\}$.

\textit{(001) well}. Out-of-plane  [001] stress keeps the 2-valley and 4-valley degeneracy and tunes the energy distance $E_{z}-E_{x}$ between them. In-plane [100](or [010]) stress breaks the 4-valley degeneracy into two groups and tunes $E_{x}-E_{y}$ while keeping the splitting $E_{y(x)}-E_{z}>0$ unchanged.

Under the $[001]$ stress, the spin relaxation keeps the unstrained form in Eq.~(\ref{eq:factor_001}) with  $E_{z}-  E_{x}$ to be tunable. In the $n=0$ limit, $d_{v_1,0;v_2,0}=2d/3$ for the square well, and about $( 9\pi \hbar)^{2/3}/[4(2\,{\rm min}[m_{001,v_1}, m_{001,v_2}] eF)^{1/3}] $ for the triangular well. In this limit we have the additional possibility that $E_{x,0}<\varepsilon_F <E_{z,0}$ which results in strong $\mathbf{s}$-anisotropic spin lifetime.

Under the $[100]$ (or $[010]$) stress, one may have $f$-process scattering available between $n=0$ subbands, by pushing $x(y)$ valleys lower towards  $z$ valleys and $y(x)$ valleys further away. This may require a relatively large compressive stress for narrow wells. In this situation, the form factor goes to $\mathcal{G}^{(001)}_{100}(\mathbf{s}) = \mathcal{S}(s_y) m_t /(1+\sqrt{m_t/m_l})$  and $s_y\leftrightarrow s_x$ for $\mathcal{G}^{(001)}_{010} (\mathbf{s})$.

\textit{(110) well}. The unique feature for (110) well under stress is that the 2-valley and 4-valley groups remain respectively degenerate, for the out-of-plane  [110] stress and all the in-plane stress directions including [001] and $[\bar{1}10]$. Therefore, the spin relaxation rate expression follow the unstrained one, only with the energy distance between $E_z$ and $E_x$ tunable by stress.

The stress dependence of 2D spin relaxation, arising entirely from the valley-dependent subband structure of the 2DEG, is a characteristic feature of the Yafet impurity process being considered in our current work, which is completely absent in the Rashba/Dresselhaus-based DP relaxation mechanism.

\section{Experimental Implications}\label{sec:experimental_implications}

In this section, we discuss the potential experimental implications of the impurity-driven spin relaxation mechanism.  Variational calculations for the 2DEG subband structure are carried out which are valid for a broader range of MOSFET parameters, including both inversion and accumulation layers, and use variables that are convenient to compare with experiments. Possible experimental proposals are discussed to differentiate  major contributions to the spin relaxation in Si 2DEG.
In particular, the few existing 2D Si spin relaxation measurements available in the literature (see our discussion below) have all been interpreted using the DP relaxation mechanism although the quantitative agreement between theory and experiment is in general not satisfactory.

Far fewer spin relaxation measurements have been made on the Si 2DEG than on the bulk Si. In n-type bulk Si, with the inversion symmetric lattice structure, Elliott-Yafet (EY) spin relaxation is the dominant mechanism for conduction electrons. It is established that the scattering is caused  mainly by the electron-phonon interaction at elevated temperatures \cite{Elliott_PR54,Yafet_SSP63, Cheng_PRL10, Li_PRL11, Tang_PRB12,Song_PRB12}, or by various processes involving impurities under high impurity density and low temperatures (see, e.g., the review in \cite{Zutic_RMP04} and Fig.~2 in \cite{Song_PRB12}).  Despite the weak SOC in Si, spin lifetime $\tau_s$ is only of the order of 10 ns at room temperature, and 0.1-100 ns at low temperatures and high donor concentrations depending on the specific donor type \cite{Shiraishi_PRB11,Ue_PRB71, Quirt_PRB72, Pifer_PRB75, Ochiai_PSS78, Zarifis_PRB98, Song_PRL14}. In comparison, in 2DEG several additional features emerge with respect to spin relaxation. Aside from the tunability of spin lifetime and anisotropy showed in Sec.~\ref{sec:detailed_results_for_2DEG}, the DP spin relaxation becomes relevant, caused by the inversion-breaking structure and interfaces and the associated Rashba/Dresselhaus field. A third feature is the large amount of interface disorder, especially in MOSFETs, which may produce valley-spin-flip scattering. As the 2DEG quantum limit is studied typically at low temperatures, the spin relaxation is determined by scattering with impurities, where EY and DP mechanisms happen respectively during and between the scattering events. In the following, we quantify the spin lifetime in the quantum limit based on our studied Yafet mechanism, taking into account the experimentally measured parameters and the uncertainty in the interface disorder.

\begin{figure}[!htbp]
\centering
\includegraphics[width=8.5cm]{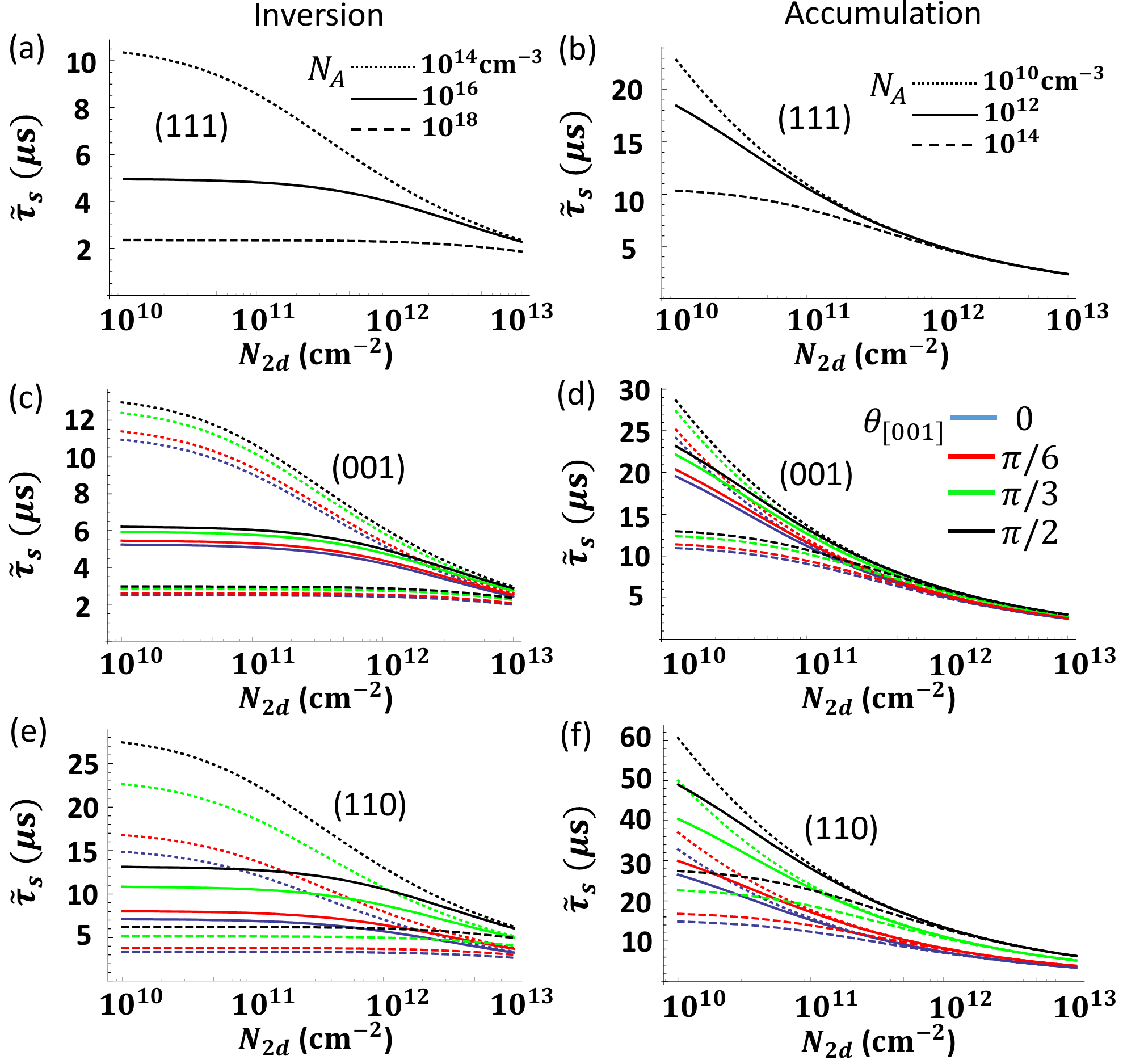}
\caption{ Normalized spin relaxation time independent of impurity density, $\tilde\tau_s= (N_i/10^{16} \textrm{cm}^{-3}) \tau_s$, for an inversion [(a), (c) and (e)] or accumulation [(b), (d) and (f)] layer in the Si MOSFET setup.  All three 2DEG plane orientations are shown: (111) in (a) and (b), (001) in (c) and (d) and (110) in (e) and (f). Only the lowest subbands are occupied [for the (001) case, the lowest subbands in the 4 higher valleys are also occupied]. They are plotted as functions of 2D electron density, $N_{2d}$, and for acceptor density (related to potential from the depletion layer), $N_A=10^{14}, 10^{16}, 10^{18}$ for the inversion layer and $10^{10}, 10^{12}, 10^{14}$ for the accumulation layer case. As previous calculations, we also give spin orientation dependence for the (001) and (110) cases.
}\label{fig:tau_s_tri_variational}
\end{figure}

To make comparison with experiments, we take the acceptor concentration $N_A$ and 2D electron density $N_{2d}$ as two independent variables. We further take the effective impurity density $N_i$ in the 2DEG as a separate variable as it can be significantly different from that in the bulk. To apply to a wide range of parameter values and to both inversion as well as accumulation layers, we relax the triangular approximation and adopt the variational subband wavefunction \cite{Fang_PRL66, Stern_PRB72,Stern_PRL74},
\begin{eqnarray}
\xi(z)= \sqrt{\frac{b^3}{2}} z \exp(-\frac{b z}{2}).
\end{eqnarray}
with a variational variable $b$. After the numerical energy minimization to find $b$ taking into account both the depletion and 2DEG layer potentials (the band bending takes 1.1 eV and 45 meV for inversion and accumulation respectively), simple expressions for the effective width $d_{v_1,0,v_2,0}$ [see Eq.~(\ref{eq:d_rev})] can be obtained in terms of  $b$:
\begin{eqnarray}
d_{v_1,0,v_2,0}= \frac{(b_1+b_2)^5}{ 6 b_1^3 b_2^3},
\end{eqnarray}
where $b_1$ and $b_2$ are respectively from valleys $v_1$ and $v_2$ which may have different $m_z$ masses. Substituting the obtained effective widths into Eqs.~(\ref{eq:U_xy}), (\ref{eq:tau_s_2d_111}), (\ref{eq:factor_111}) and (\ref{eq:factor_001}), one can obtain $\tau_s$ for three 2DEG orientations, (111), (001) and (110), for any arbitrary spin orientation, as a function of $N_A$, $N_{2d}$ and $N_i$. We collect all useful information in Fig.~\ref{fig:tau_s_tri_variational}. For visual clarity, we leave out the reciprocal linear dependence on $N_i$ and plot $\tilde{\tau}_s = \frac{N_i}{10^{16} \textrm{cm}^{-3}} \tau_s$. For the inversion (accumulation) layer case, we choose the majority (minority) acceptor density $N_A$ as $10^{14}, 10^{16}$ and $10^{18}$ cm$^{-3}$ ($10^{10}, 10^{12}$ and $10^{14}$ cm$^{-3}$), and $10^{10}<N_{2d}< 10^{13}$ cm$^{-2}$, covering typical experimental choices.  As expected, $\tilde\tau_s$ decreases slightly with increasing $N_{2d}$ due to the steeper confinement. Similar trends occur with $N_{A}$. More importantly, of course, the absolute time $\tau_s\propto \tilde{\tau}_s/N_i$ decreases much faster with $N_i$. For clean interfaces, $N_i\approx N_A(N_D)$ for the inversion (accumulation) layer. However, for some highly disordered Si/SiO$_2$ interfaces, including oxide charges on the SiO$_2$ side, $N_i$ may be much larger than the majority dopant density. In the high $N_{2d}$ limit, the potential confinement is dominated by $N_{2d}$ over $N_{\rm depl}\propto \sqrt{N_A}$ \cite{Ando_RMP82}, and as a result $\tilde\tau_s$ converges for different $N_A$'s.

Finally, we should note that the SOC parameter $\Delta_{\rm so}$ used here is $0.1$ meV, the value for As dopants. It is $0.03$ or $0.3$ meV for P or Sb \cite{Castner_PR67}. For interface disorder or majority acceptors, $\Delta_{\rm so}$ and $\eta$ [see Eq.~(\ref{eq:U_s_free})] need to be studied separately. This can be an important study in the future, and we will come back to this issue at the end. In fact, the effective SOC parameters of the interface impurities are an important unknown in the theory, which can be adjusted to get agreement between our theory and all existing experimental data.  We refrain from doing so, however, emphasizing that if the measured 2D spin relaxation time shows a positive correlation with the quality of the interface (i.e. improving interface quality leads to longer spin relaxation time), then it is likely that the impurity induced Yafet mechanism discussed in this paper is playing a dominant role in contrast to the DP mechanism which mostly leads to a lower spin relaxation time with higher mobility.

Now we briefly discuss some available experimentally measured $T_1$($\equiv \tau_s$ in our notation) in Si 2DEG. We stress that these samples are not particularly highly doped and therefore our mechanism is not expected to be dominant unless extrinsic disorder associated with interface impurities are playing a crucial role. Reference~\cite{Tyryshkin_PRL05} measured $T_1$ in Si/SiGe (001) quantum wells with relative high quality interface and mobility $\mu$. Their device I with $\mu=9$ m$^2$/Vs, 2D electron density $N_{2d}=3\times 10^{11}$ cm$^{-2}$ and well thickness $d=20$ nm, yields a $T_1=2.0 \,\mu$s. Their device II with $\mu=19$ m$^2$/Vs, $N_{2d}=1.7\times 10^{11}$ cm$^{-2}$ and  $d=15$ nm, yields a $T_1=0.95 \,\mu$s. These $N_{2d}$ and $d$ combinations indicate that only the ground subbands in the $\pm z$ valley are likely occupied.

We show in the following that our mechanism cannot quantitatively account for the measured $T_1$, even if we assume the lowest subbands in the $\pm x$ and $\pm y$ valleys are occupied.    Using our calculation that leads to Fig.~\ref{fig:tau_s_001surf_sq}, one needs a 3D impurity density $N_i\approx 3.3\times 10^{16}$ cm$^{-3}$ or an effective 2D impurity density $n_i\approx N_i d\approx 6.6\times 10^{10}$ cm$^{-2}$ for device I, and $N_i\approx 5\times 10^{16}$ cm$^{-3}$ or $n_i\approx N_i d\approx 7.5\times 10^{10}$ cm$^{-2}$ for device II. The precise $n_i$ depends weakly on the detailed impurity distribution in the 2DEG.  To estimate the experimental impurity density residing in the quantum well, we use the theoretical result of Ref.~\cite{Hwang_PRB13} which relates mobility with the charged impurity. From Fig.~1 of that paper, for the (001) well orientation with two ground valleys, one needs $n_i\approx 10^{10}$ cm$^{-2}$ for device I and $n_i\approx 4\times 10^9$ cm$^{-2}$ for device II. These $n_i$'s make our spin relaxation mechanism too weak to yield the measured $T_1$ time. The spin anisotropy in our calculation has the same sign as in the measurement but not as large in magnitude [$T_1(\theta=\pi/2)/T_1(\theta=0)=1.1$ versus measured 1.5].

We do, however, mention that our mechanism using these estimated experimental impurity densities gives $T_1$ values within an order of magnitude of the measured $T_1$ values. Given the uncertainties associated with the impurity SOC parameters, the possibility that the Yafet mechanism is perhaps playing a (minor) role in the experiment cannot be ruled out although it does appear that the main spin relaxation mechanism  in these high mobility Si/SiGe
quantum wells is likely to be the DP mechanism.

Reference~\cite{Shankar_PRB10} measured $T_1$ time for 2DEG in a Si/SiO$_2$ (001) accumulation layer doped with $10^{14}$ P donors. At the gate voltage of 2 V, $T_1=0.33\, \mu$s while $\mu=1$ m$^2$/Vs and $N_{2d}=4\times 10^{11}$ cm$^{-2}$ \cite{Shankar_thesis10}. Once again we check the effect of our mechanism by assuming for a moment that the $\pm x$ and $\pm y$ valleys are reached. From Fig.~\ref{fig:tau_s_tri_variational}, one needs about $N_i\approx 3\times 10^{17}$ cm$^{-3}$ or $n_i\approx 4\times 10^{11}$ cm$^{-2}$. Note the unknown $N_A$ value affects the result only slightly (a factor less than 2). From the measured mobility at 5 K, we can deduce $n_i\approx 10^{11}$ cm$^{-2}$ \cite{Hwang_PRB13}. As a result, the impurity density is again too small to induce the measured $T_1$ time by our mechanism, even if the finite mobility is entirely caused by impurities in the 2DEG region and $\Delta_{\rm so}=0.1$ meV. But now our mechanism gives a $T_1$ which is within a factor of 4 of the measured value, indicating that for spin relaxation in disordered Si MOSFETs, perhaps our impurity-driven mechanism is playing a more important quantitative role.  This is not unexpected since Si MOSFETs typically have larger impurity densities than Si/SiGe quantum wells, leading to possibly stronger spin relaxation due to the Yafet mechanism.  Note that the DP mechanism does not find agreement with the experimental data either, which can be verified by the calculation in Ref.~\cite{Tahan_PRB05} in combination with the experimental parameters.  It is possible that in Si MOSFETs both DP and Yafet mechanisms are operational in producing the observed low value of $T_1$ in the experiment.  Obviously, more experimental measurements are essential in understanding this important puzzle.

We propose several experimental ways to properly investigate the nature of spin relaxation in Si 2DEG. To begin with, (111) and (110) orientations are better suited for our intervalley spin-flip mechanism to have important contribution, as we have mentioned before. Moreover, a lowering of the $x$ and $y$ valleys in the (001) 2DEG may also show a sudden jump of spin relaxation rate which serves as a turn-on signal of our mechanism. This valley tuning can be achieved by external stress or gating, as emphasized in Sec.~\ref{sec:detailed_results_for_2DEG}. A similarly sudden change in $\tau_s$ anisotropy can also occur for the (110) 2DEG due to our mechanism.

Apart from the 2DEG plane orientation, a number of aspects are important in the experimental verification of our proposed spin relaxation mechanism.  First, it is crucial (and we urge future experiments) to measure a series of samples with different mobilities ($\mu$) at same carrier densities (and all other parameters). The (anti)correlation of $\tau_s$ with $\mu$ is a characteristic signature for Yafet (DP) spin relaxation mechanism \cite{Zutic_RMP04}.  The crossover occurs at modest doping levels, as the spin lifetime from the DP mechanism rises rapidly past tens of $\mu$s already around mobility $5\times 10^4$ cm$^2/$Vs \cite{Tahan_PRB05}. In particular, our mechanism should become dominant when the 2DEG region is heavily doped.

Second, there is an  positive correlation between conduction electron density and spin relaxation in the DP mechanism. The Rashba or generalized Dresselhaus field scales linearly with the wavevector ($\mathbf{k}$) measured from valley bottom yet our spin flip matrix elements depend little on $\mathbf{k}$.
Third, for the MOSFET setup, it is useful to measure $\tau_s$ separately for both the bulk Si and the 2DEG to deduce the contribution of interface disorder to the 2DEG spin relaxation.

It is also possible to deduce the distribution of 2DEG impurities from the gate voltage dependence of $\tau_s$: from Eqs.~(\ref{eq:tau_s_2d}) and (\ref{eq:U_2d}), $\tau_s\propto d$ for a uniformly distributed $N_i(z)$ while $\tau_s\propto d^2$ if all impurities are concentrated at the interface [$N_i(z)\propto\delta(z)$].
For the Si/SiGe setup, making two-sided symmetric confinement may separate out the contribution from the DP spin relaxation, as the change of interfacial symmetry property greatly affects the DP mechanism through the envelope functions but leaves the Yafet one the same.
In addition, as our calculated spin orientation dependence [rooted in Eqs.~(\ref{eq:U_xy})-(\ref{eq:U_zy})] is distinct from that of the DP mechanism due to the Rashba or Dresselhaus field, $\tau_s$ anisotropy measurement can also help to disentangle the two contributions (for the large magnetic field limit, we note that, the DP mechanism is partially suppressed similar to the bulk case \cite{Zutic_RMP04,Wilamowski_PRB04,Glazov_PRB04}).

Parenthetically, while this work does not focus on the DP mechanism, we point out the existing studies concerning its various contributions \cite{Ganichev_PSS14}.  Different views have emerged to account for the same experimental measurement in Si/SiGe quantum well \cite{Wilamowski_PRB02}, being it dominated by the Rashba field \cite{Wilamowski_PRB02,Wilamowski_PRB04,Tahan_PRB05} or the Dresselhaus one \cite{Nestoklon_PRB08, Prada_NJP11}. They lead to different SOC anisotropy but similar overall spin relaxation rate as both SOC fields scale linearly in wavevector in the 2DEGs. DP spin relaxation and its anisotropy has also been studied in Si/SiGe quantum dots \cite{Raith_PRB11}. Up to the present, the relative magnitude of the Rashba and Dresselhuas-like SOC has yet to be verified experimentally \cite{Tahan_PRB14}.

Finally, our mechanism relies on the short-range interaction with the impurity core and directly measures  the SOC strength of the impurity atoms. The spin relaxation rate scales quadratically with $\Delta_{\rm so}$ [see Eqs.~(\ref{eq:U_2d}) and (\ref{eq:tau_s_2d})], and it increases significantly by switching from low atomic-number to high atomic-number dopants for the same density. Therefore,  different types of impurities  that lead to similar mobility may yield very different $\tau_s$ times according to their SOC strengths,  a unique signature of this spin relaxation process.

We suggest future spin relaxation measurements in 2D Si systems as a systematic function of mobility, carrier density, impurity type, surface and spin orientation, and applied stress in order to develop a complete understanding of the mechanisms controlling spin relaxation of free carriers near Si surfaces.  The few existing measurements simply do not have enough information for a definitive conclusion.

Last, in order to establish the relative strength of our spin relaxation rate in comparison to the momentum relaxation rate,   which determines the device charge mobility, we calculate their relative ratio ($\nu$) for a few representative cases where impurities are the dominant source of scattering (i.e. at low temperatures where phonons are unimportant). We take a simplified uniform distribution of the highly doped 2DEG. Define $\nu=\tau^{2d}_m/\tau^{2d}_s$, where $\tau^{2d}_s$ follows from our Eq.~(\ref{eq:tau_s_2d}) and the momentum relaxation time $\tau^{2d}_m$ takes the form appropriate for mobility calculations in 2D transport studies \cite{Hwang_PRB13,DasSarma_PRB13}.

The momentum scattering matrix elements are governed by the well-known screened Coulomb interaction in the intravalley scattering, as appropriate for scattering by the random charged impurities. For this interaction, the impurity distribution profile can be approximated as a $\delta$ function  normal to the 2DEG plane. Under the 2D RPA screening, the momentum relaxation rate in the quantum limit is given as \cite{Hwang_PRB13,DasSarma_PRB13},
\begin{eqnarray}\label{eq:momentum_rate}
\frac{1}{\tau^{2d}_m} = \frac{4\pi e^4 m_{\rm eff} n_i}{\hbar^3 \kappa^2 k_F^2}
\int^1_0  \frac{dx x^2}{(x + q_{\rm TF}/2k_F)^2 \sqrt{1-x^2}},
\end{eqnarray}
where $m_{\rm eff}$ is the conductivity effective mass different for each specific 2DEG orientation \cite{Ando_RMP82,Hwang_PRB13}, $n_i$ is the 2D impurity density, permittivity $\kappa$ is the Si permittivity,  $k_F$ is the 2D Fermi wave number, 2D Thomas-Fermi wave number $q_{\rm TF}=m_{\rm eff} e^2 g/\hbar^2 \kappa$, and $g=g_v g_s$ is the number of populated valleys including the spin degree of freedom ($v$ and $s$ denoting valley and spin).

\begin{figure}[!htbp]
\centering
\includegraphics[width=7.5cm]{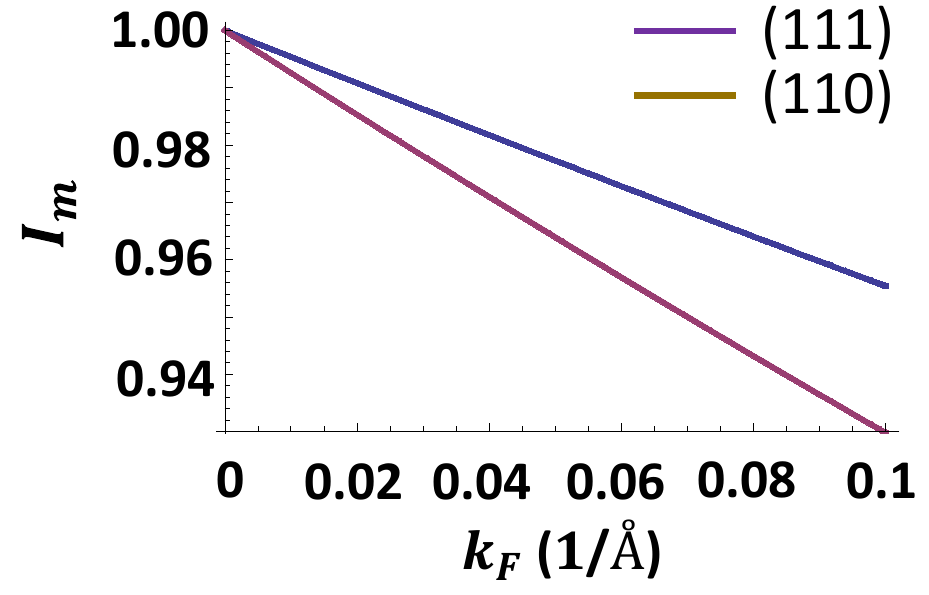}
\caption{The $k_F$-dependent factor in momentum scattering rate [Eq.~(\ref{eq:momentum_rate})], $I_m \equiv \frac{1}{\pi} (\frac{k_F}{q_{\rm TF}})^2 \int^1_0  \frac{dx x^2}{(x + q_{\rm TF}/2k_F)^2 \sqrt{1-x^2}}$, for (111) and (110) 2DEG orientations respectively, over $0< k_F<0.1 \AA^{-1}$.
}\label{fig:momentum_integral}
\end{figure}

The dependence on Fermi level (similarly, on $k_F$ or $N_{2d}$) is very slow for both $\tau^{2d}_s$ and $\tau^{2d}_m$ within a given Fermi energy window between 2D subbands \cite{DasSarma_PRB13}. This can be clearly  seen for $\tau^{2d}_s$ over many order of magnitudes of $N_{2d}$ from Fig.~\ref{fig:tau_s_tri_variational}. For $\tau^{2d}_m$, we plot its explicit dependence on $k_F$ in Fig.~\ref{fig:momentum_integral} for (111) and (110) 2DEG orientations for which both leading-order momentum and spin relaxation rates are nonvanishing in the quantum limit. They are both nearly constant over the large region $k_F\leq 0.1 \AA^{-1}$, i.e., about 10\% of the length of the Brillouin zone.

This near independence of Fermi level allows us to obtain simple estimation for $\nu$ for each specific 2DEG orientation. To get the leading-order estimate, it is sufficient to substitute $\tau^{(111)}_s$ and $\tau^{(110)}_s$ with the square well results through Eqs.~(\ref{eq:tau_s_2d_111}), (\ref{eq:factor_111_sq}) and a variation of (\ref{eq:factor_001_sq}). We have the following results,
\begin{eqnarray}
\nu_{111} \!&\!= \!&\!\frac{400 a_B^6\Delta_{\rm so}^2 }{\hbar^4 d^2} \sqrt{\frac{m_t(m_t\!+\! 2m_l)}{3}} \!\big/\!\left(\frac{1}{m_t}\! + \!\frac{3}{m_t\! +\! 2 m_l}\right),\quad
\\
\nu_{110} \!&\!=\!& \!\frac{128 a_B^6\Delta_{\rm so}^2 }{\hbar^4 d^2} \sqrt{m_t m_l} \big/\left(\frac{1}{m_t} + \frac{2}{m_t + m_l}\right),
\end{eqnarray}
for (111) and (110) 2DEGs respectively. With respect to typical $\Delta_{\rm so}$ and $d$ parameters, $\nu_{111}=1.5\times 10^{-5} (\frac{\Delta_{\rm so} }{0.1 {\rm meV}})^2(\frac{20{\rm nm}}{d} )^2$ and $\nu_{110}=5.6\times 10^{-6} (\frac{\Delta_{\rm so} }{0.1 {\rm meV}})^2(\frac{20{\rm nm}}{d} )^2$. Obviously, this ratio depends quadratically on the impurity SOC constant $\Delta_{\rm so} $. For the expected typical values of the impurity SOC $\Delta_{\rm so}$ in Si and the 2DEG width $d$, this ratio $\nu$ varies between $10^{-6}$ and $10^{-4}$. In comparison, we note that for intrinsic phonon-induced spin and momentum relaxation rates in 3D bulk Si, this ratio is around $10^{-5}$ \cite{Cheng_PRL10}, which is determined completely by the host Si SOC.

\section{Summary and Outlook}\label{sec:summary}

We have introduced in Si 2DEG a previously overlooked yet important spin relaxation mechanism due to electron-impurity scattering. This mechanism dominates over other spin relaxations in the multi-valley Si conduction band as impurity density increases, and can be significantly suppressed when electrons are transferred into two opposite ground valleys by specific 2DEG orientations and stress configurations. We provide the general expression for obtaining the leading-order spin relaxation rate under arbitrary confinement potential, applied stress,  and subband occupation. We calculate quantitatively the ($T_1$) spin relaxation time $\tau_s (\mathbf{s})$ as a function of spin orientation $\mathbf{s}$, as well as of the conduction electron density and confinement strength for the representative square and triangular wells.

Moreover, the consequences of various stress configurations have been worked out in details. Importantly, this newly discovered spin relaxation mechanism combined with the Si 2DEG setup provides interesting possibilities to tune spin lifetime as well as its dependence on spin orientation (or applied  magnetic field direction) substantially by on-chip gate voltages and possibly by local stress. Such a tunability of spin relaxation in MOSFET-type Si devices could have potential spintronic applicability.

Also crucially, we provide experimental ways (elaborated in Sec.~\ref{sec:experimental_implications}) to verify our spin relaxation mechanism and distinguish it from the DP spin relaxation effect from the generalized Rashba/Dresselhaus field in Si 2DEGs, by exploiting their different dependence on impurity densities and types, on the interface symmetry properties, and on 2DEG plane, spin and stress orientations.

Regarding a general expansion of this model, we point out that for 2DEG near the interface with considerable amount of disorder, a variation of our impurity-driven intervalley spin-flip process may become quantitatively important in determining the spin relaxation rate. As  mentioned in the introduction, DP spin relaxation mechanism alone leads to much longer spin lifetime for low-mobility 2DEG than observed experimentally \cite{Tahan_PRB05}. However, spin lifetime is apparently shorter in 2DEG near typical Si/SiO$_2$ interfaces, indicating  impurity-driven Elliott-Yafet spin relaxation.  While our spin-flip matrix elements [Eq.~(\ref{eq:U_s_free})] apply specifically to substitutional impurities in Si with their given symmetry, it is a basic rule that lower-symmetry disorder inherits the allowed transition matrix elements. Thus the key idea of zeroth-order intervalley spin-flip scattering \cite{Song_PRL14} robustly holds for irregular defects, with additional scattering channels potentially open depending on the specific defects.  It is therefore possible that interface impurities (even when they are completely nonmagnetic as our theory entirely restricts itself to-- any magnetic interface impurities will of course very strongly affect spin relaxation near the surface through direct magnetic spin-flip scattering) are playing a strong role in determining the 2D spin relaxation time in disordered Si/SiO$_2$ MOSFETs by participating in the Yafet process identified and analyzed in the current work.  Obviously, figuring this out remains an open and important future experimental challenge in Si spintronics.

This work is supported by LPS-MPO-CMTC.

\appendix
\section{physics of Intervalley coupling in Si and symmetry analysis}\label{app}

To be self-contained, we provide the essential physical picture of intervalley coupling in bulk Si and the relevant symmetry analysis and selection rules for Sec.~\ref{sec:theoretical formulation}.

\begin{figure}[!htbp]
\centering
\includegraphics[width=7.5cm]{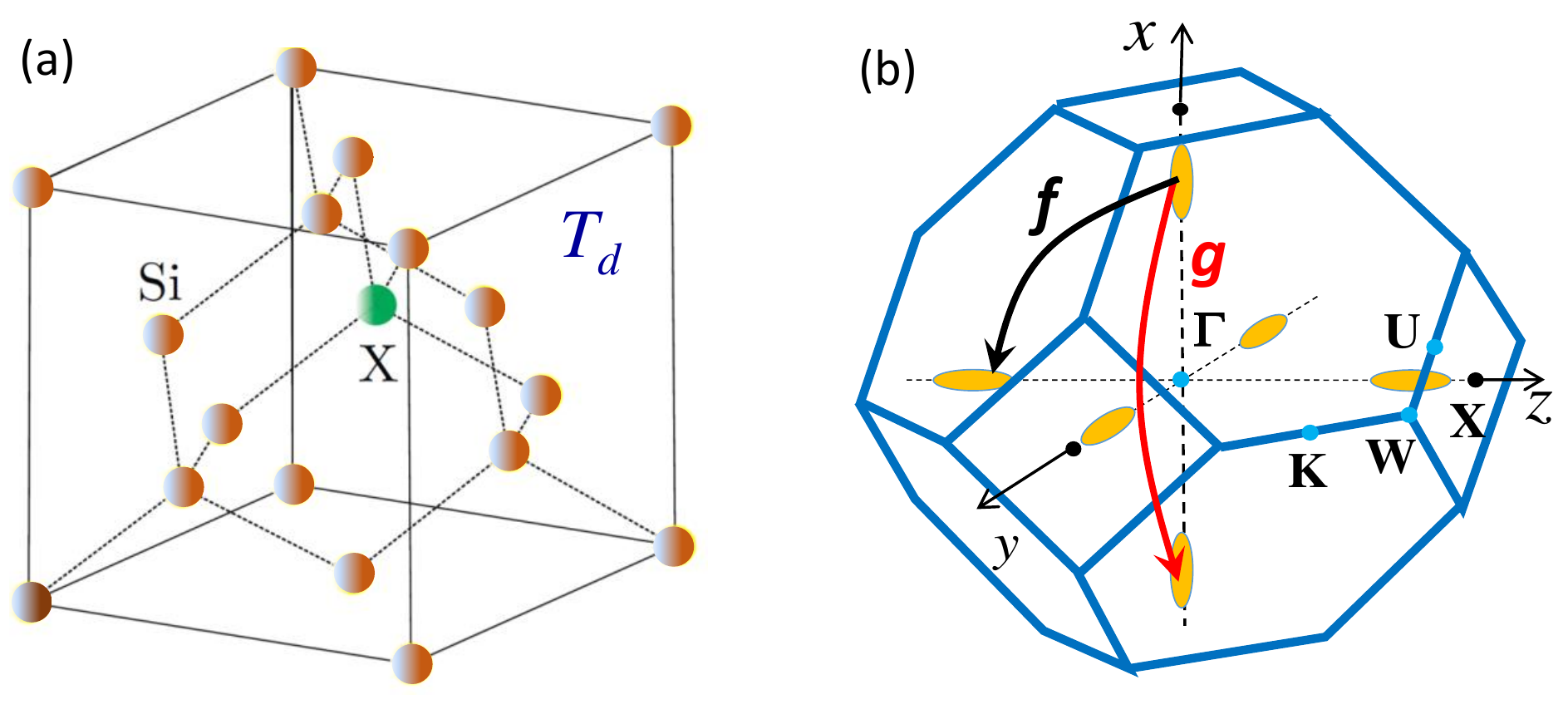}
\caption{(a) The Si crystal lattice, with one of its Si atoms replaced by an impurity denoted as ``X''. As a result, the symmetry of the whole Hamiltonian system is reduced to that of the $T_d$ point group. (b) The Brillouin zone of the Si crystal. The yellow ellipsoids mark the low-energy surface of the 6 conduction valleys. Two representative examples are marked for intervalley $g$ and $f$ processes.
}\label{fig:app}
\end{figure}

Bulk Si has the crystal structure [Fig.~\ref{fig:app}(a) in absence of the impurity substitution] consisting of two sets of interpenetrating face center cubic lattices, and a space symmetry group $O^7_h$. Its lowest conduction band in the wavevector space has its bottoms not at the center of the Brillouin zone but along the cubic axes directions. Crystal symmetry determines that 6 energy valleys reside cylindrically along $\pm x, \pm y$ and $\pm z$ axes. This well-known multivalley picture of Si supplies relevant information for the electron states involved in this work. The transitions between these electron states residing near the bottom of the conduction valleys can obviously be classified into three groups [see Fig.~\ref{fig:app}(b)]: (I) within the same valley (``intravalley''), (II) between two opposite valleys (``intervalley $g$ process''), and (III) between two non-opposite valleys (``intervalley $f$ process'').

The particular scattering potential we deal with comes from the impurity which replaces one of the Si atoms. This impurity immediately invalidates the translational symmetry of the Si crystal, and as a result the symmetry of the Hamiltonian system falls into a point group around the impurity [Fig.~\ref{fig:app}(a)]. This point group has the same symmetry operations as a tetrahedron molecule: $C_2$ rotation about $x,y$ or $z$ axis, $C_3$ rotation about body diagonals, $\sigma$ reflection about the face diagonal planes, and $S_4$ ($C_4$ followed by reflection) about $x,y$ or $z$ axis, and is called the $T_d$ group.

To utilize the symmetry property of this system for selection rules, we work with symmetrized electrons states  by linearly combining 6 different valleys rather than states in each individual valley as one is used to. The 6 combinations are as follows \cite{Kohn_SSP57}
\begin{eqnarray}
\psi_{A_1}&=& \frac{1}{\sqrt{6}}(1, 1, 1, 1, 1, 1);  \\
\psi_{E^I}&=&\frac{1}{2}(1, 1, -1, -1, 0, 0),\\
\psi_{E^{I\!I}}&=&\frac{1}{2\sqrt{3}}(1, 1, 1, 1, -2, -2);\\
\psi_{T_2^{I}}&=&\frac{1}{\sqrt{2}}(1, -1, 0, 0, 0, 0),\\
\psi_{T_2^{I\!I}}&=&\frac{1}{\sqrt{2}}( 0, 0,1, -1, 0, 0),\\
\psi_{T_2^{I\!I\!I}}&=&\frac{1}{\sqrt{2}}( 0, 0, 0, 0,1, -1);
\end{eqnarray}
where the ordering of the 6 components of the state vectors is the valley bottom state along $+x$, $-x$, $+y$, $-y$, $+z$ or $-z$ axis, respectively. Each new state is given a name at the subscript of $\psi$, following the well-established naming system (see the $T_d$ group character table in \cite{Bradley_Cracknell72} or \cite{Song_PRL14}). The selection rules immediately follow, since only the same-symmetry states can couple while different-symmetry states are not mixed by the scatterer potential which transforms as the identity in this group. Once we get the scattering matrix elements that do not vanish, we can easily make linear combinations between them to transfer back to the familiar intravalley and intervalley $g$ and $f$ processes \cite{Song_PRL14}.

Thus far, we have not considered spin degrees of freedom or SOC. To include spin, we can expand the basis to be the product space of 6 valleys and 2 spins. It turns out two $\bar{F}$ states emerge from this valley-spin coupling. To be concrete, the multiplication expressions are as follows. The pure spin transforms as $\bar{E}_1$, and then we have
\begin{eqnarray}
A_1 \times \bar{E}_1 &=& \bar{E}_1, \label{eq:A1_SOC}\\
E \times \bar{E}_1 &=& \bar{F},\label{eq:E_SOC}\\
T_2 \times \bar{E}_1 &=& \bar{E}_2 + \bar{F} . \label{eq:T2_SOC}
\end{eqnarray}
We may follow a similar procedure as the spinless case to obtain spin-dependent scattering selection rules \cite{Song_PRL14}. Only states with the same symmetry can be coupled. Among all the 5 nonvanishing couplings [each of the 4 states in Eqs.~(\ref{eq:A1_SOC})-(\ref{eq:T2_SOC}) coupling to itself, as well as the inter-coupling of the two $\bar{F}$ states from Eqs.~(\ref{eq:E_SOC}) and (\ref{eq:T2_SOC})], we find that there are spin-flip terms in two of them: the difference between $\bar{E}_2$ and $\bar{F}$ self-coupling matrix elements from  Eqs.~(\ref{eq:T2_SOC}), and the inter-coupling matrix element between two different $\bar{F}$. That leads to the two terms in Eq.~(\ref{eq:U_s_free}), respectively. After transforming back to the intravalley, intervalley $g$ and $f$ processes, we find \cite{Song_PRL14} both terms contribute to the $f$-process spin flip.

\end{document}